\documentclass{article}
\usepackage{iclr2024_conference}
\iclrfinalcopy
\usepackage{multirow}
\usepackage{color}
\usepackage{lineno}
\usepackage{natbib}

\usepackage{hyperref}
\usepackage[utf8]{inputenc}
\usepackage{amsmath,amsfonts,amssymb}
\usepackage[T1]{fontenc}    % use 8-bit T1 fonts
\usepackage{url}            % simple URL typesetting
\usepackage{booktabs}       % professional-quality tables
\usepackage{nicefrac}       % compact symbols for 1/2, etc.
\usepackage{microtype}      % microtypography
\usepackage{graphicx}
\usepackage{adjustbox}

\usepackage{color}
\definecolor{dkgreen}{rgb}{0,0.6,0}
\definecolor{dkdkgreen}{rgb}{0,0.3,0}
\definecolor{gray}{rgb}{0.5,0.5,0.5}
\definecolor{mauve}{rgb}{0.58,0,0.82}

\usepackage{algorithm}
\usepackage{algpseudocode}
  \DeclareGraphicsExtensions{.png,.pdf}
\usepackage[symbol]{footmisc}
\usepackage{authblk}
\usepackage{physics}
\usepackage{multirow}
\usepackage{longtable}
\usepackage{upquote}
\usepackage{subfig}

\renewcommand{\thefootnote}{\fnsymbol{footnote}}

\usepackage{appendix}

\usepackage{listings}

\usepackage{dsfont}

\newcommand{\mbf}[1]{\mathbf{#1}}

\newcommand{\R}{\mathbb{R}}
\newcommand{\C}{\mathbb{C}}
\newcommand{\N}{\mathbb{N}}
\newcommand{\Z}{\mathbb{Z}}
\newcommand{\F}{\mathbb{F}}
\newcommand{\Q}{\mathbb{Q}}

\newcommand{\x}{\mbf{x}}

\usepackage{graphicx} % Required for inserting images

\begin{document}

\title{Preliminary Report on \textit{Mantis Shrimp}: a Multi-Survey Computer Vision Photometric Redshift Model}

\author{%
  Andrew W. Engel$^{1}$, Gautham Narayan$^{2,3,4}$, and Nell Byler$^{1}$\\
  $^{1}$Pacific Northwest National Laboratory, 902 Battelle Blvd, Richland, WA 99354\\
  $^{2}$Department of Astronomy, University of Illinois at Urbana-Champaign, Urbana, IL 61801, USA \\
  $^{3}$Center for AstroPhysical Surveys, National Center for Supercomputing Applications, Urbana, IL 61801, USA 
  $^{4}$Illinois Center for Advanced Studies of the Universe, University of Illinois Urbana-Champaign, Urbana, IL 61801, USA 
  \texttt{\{andrew.engel,eleanor.byler\}@pnnl}\\
  \texttt{gsn@illinois.edu}\\
}

\date{February 2023}

\maketitle

\begin{abstract} The availability of large, public, multi-modal astronomical datasets presents an opportunity to execute novel research that straddles the line between science of AI and science of astronomy. Photometric redshift estimation is a well-established subfield of astronomy. Prior works show that computer vision models typically outperform catalog-based models, but these models face additional complexities when incorporating images from more than one instrument or sensor. In this report, we detail our progress creating \textit{Mantis Shrimp}, a multi-survey computer vision model for photometric redshift estimation that fuses ultra-violet (GALEX), optical (PanSTARRS), and infrared (UnWISE) imagery. We use deep learning interpretability diagnostics to measure how the model leverages information from the different inputs. We reason about the behavior of the CNNs from the interpretability metrics, specifically framing the result in terms of physically-grounded knowledge of galaxy properties. 
\end{abstract}

\section{Introduction}
%What is a photometric redshift and why are they important?
The ability to measure a redshift to a far-away galaxy is fundamental to most of extragalactic astronomy, and is often used as a proxy for distance. Redshift measurement traditionally requires expensive and time-consuming spectroscopic observations; however, photometric redshift (photo-z) estimation uses broad-band image observations instead, which are less accurate but are fast to obtain and can be applied to many more objects. Ongoing surveys \citep{KIDS_plus_VIKING_BPZ,DES_Photometric_Redshifts,RedshiftsCFHS} and future surveys \citep{LSST_ScienceBook2,eulid,RomanDoc} require accurate photo-zs to complete their science objectives. Despite the importance of photo-zs, they are still the leading source of uncertainty in many cosmological analyses \citep{WeakLensingReview18}. See \citet{NewmanReview} for a recent review.  

%What is novel in they way we approach the task?
Historically, empirical photo-z models are dedicated to observations from a single instrument and rely on catalog-based model inputs \citep{Beck2016LLNSDSS,ANNzCollister2003}. However, recent research has shown that photo-z models using computer vision approaches (i.e., images as inputs) outperform catalog-based models\citep[][P19]{Pasquet2019}. Separately, prior work has demonstrated that models can successfully combine observations across multiple instruments \citep[e.g.,][]{WISE-PS1-STRMBeck22, KIDS_plus_VIKING_BPZ}, but only in the context of catalog-based photo-z models. This work bridges this gap by developing a computer vision photo-z model that fuses images from multiple instruments: ultraviolet (UV), optical, and infrared (IR). Successfully training multi-modal deep learning models is challenging; explaining modality utilization is an active area of research \citep{UniModal_Collapse}. This work provides a unique opportunity to analyze the efficacy of well-known machine learning (ML) interpretability metrics: the wavelength bands covered by each of the instruments probe distinct physical processes within a galaxy, linking metric changes to ground truth understanding. Our results are as follows:
\begin{itemize}
\item We find that the photo-z model prioritizes information from the three instruments differently.% and that the prioritization changes as a function of redshift.%, which is inline with our understanding of the underlying physical phenomena. 
We suggest that the variation seen in the interpretability metrics is primarily driven by meaningful physical processes: from underlying behavior in the spectral energy distribution of galaxies and from the surface-brightness and redshift (distance) relationship.
%\item Availability of the optical and IR bands generally causes redshift estimates to move to lower and higher values, respectively. We suggest this is due to the movement of known phenomena in the underlying spectral energy distribution of galaxies.
%\item  For the 5 bands observed by the optical instrument, we find that interpretability metrics predict a shift in band prioritization that matches a well-known reliance on a key spectral feature. 
%Above: I think this is exactly not observed.
\item We do not find compelling evidence that the model prioritizes information from the UV bands, suggesting it could be removed without penalizing performance.
\end{itemize}

\section{Methodology}
\textbf{Models:}
%point out the only odd thing about this section.
We use a ResNet50 \citep{ResNetHe2015} architecture with the input layer modified to accommodate 9-band images. We recast the numerical redshift estimation task as a classification problem, outputting a $C=200$ dimensional unit simplex vector where the $c$-th entry represents the probability that the true redshift resides in a small bin with width $\delta c$=0.005 in the span of redshifts from 0.0 to 1.0. In addition to the 9-band image input, we also include the extinction due to dust measured along line of sight for each galaxy. Following P19, the dust extinction is concatenated into the dense head of the ResNet50.

The model is trained from randomly initialized weights and we use the Adam optimizer to minimize cross entropy loss \citep{AdamOptimizerKingma2014}. A full description of the models is given in appendix~\ref{appendix:neural_networks}. We train our model for 85 epochs, stopping after an estimated 10 hours of total training time on a single V-100 GPU on a NVIDIA DGX-2 compute node. Additional details of the training process including hyperparameters and loss history visualizations are available in appendix~\ref{appendix:training}.
%We will actually take on the photo-z density estimation task \cite{ANNz2_and4000A_break_Affects_PZ}. Following the works of P19, our neural network  In effect, we will interpret the output of our model as an estimate of the posterior distribution of redshift conditioned on the data. We leave a formal definition of our task and formal definition of neural networks to .

\textbf{Data:} %(N=4246280) 
%what is most important to the reader? everything else -> appendix.
The data consists of 9-band images paired with ground-truth spectroscopic redshifts. The dataset includes $N=4.2\times10^6$ galaxies compiled from a diverse set of spectroscopic surveys \citep{SDSS_four,DESI_EDAspectroscopy, DEEP2, DEEP3, GAMA_DR2, VVDS_final_release, VIPERS_final, 6dFGS_final, WiggleZ_final}. Details on the spectroscopic surveys are provided in appendix~\ref{appendix:spectroscopic} and image data collection details are provided in appendix~\ref{appendix:photometry_download_and_description}.

For each datapoint, we query public APIs of image cutout servers for each of the three instruments, centered on the coordinates provided by the spectroscopic surveys. We include 2 UV bands from GALEX \citep{GalexSurveyPaper}, 5 optical bands from Pan-STARRS \cite{PSSurveyPaper}, and 2 IR bands from UnWISE \citep{WISE_og,UnWISE_og}. We describe the image pre-processing and augmentation pipeline in detail in appendix~\ref{appendix:photometry_processing}. In short: we ensure our images are on a quasi-logarithmic flux scale, then re-normalize images so that the pixels are between [0,1]. The different instruments have different spatial resolutions; to stack the images into a single input tensor, we resample the UV and IR bands to match the resolution of the optical bands, using nearest-neighbor interpolation. Initial experimentation shows this approach outperforms an architecture with separate instrument encoders and concatenated feature vectors.

%Spectroscopic redshifts are the most secure method of identifying redshift, typically done through monitoring emissions and absorption lines in measured spectra and comparing the wavelengths of the observed lines to their value in the rest frame. Even though spectroscopy is secure, we still ensure quality cuts to retain a high-confidence sample of target redshifts. 

%$N_{\text{eff}}$ = 253954 samples
For this preliminary report, we limit our sample to the redshift range $\in [0,1.0]$ and randomly sample 7\% of the total dataset, producing $N_{\text{eff}}$ = $2.5\times10^5$ datapoints. This ensures we can load our dataset into volatile memory at the start of training to avoid the IO bottleneck of loading individual batches throughout training, which allows us to be more agile in this preliminary investigation. Our sample is split into a standard train-val-test split with 65\%-20\%-15\% of the original $N_{\text{eff}}$ samples. 

%Images are downloaded in an original linear-flux scale, so we pre-process our data by converting each image into a luptitude scale using the arcsinh scaler, which avoids issues of negative pixel values of the more standard logarithmic-flux scale (Beck20). Images are rescaled to achieve a survey-specific maximum pixel value of $\approx$ 1, such that most of the sample  have pixel values distributed between [0,1]. A visualization of example images are provided in appendix~\ref{appendix:image_examples}. We additionally employ data augmentation with random flips and rotations of the images, but crucially we do not shift the centers of the image. We crop around the centers of images so that each spans a constant 30 arcsec width and height. We had queried each API so that this crop removes any undefined pixels from the rotation. finally, because the pixel resolution differs for each survey, we upsample the GALEX and UnWISE images to the PanSTARRS $(120 \times 120)$ pixel grid for early fusion experiments.

\textbf{Evaluation Metrics:}
\label{sec:metrics}
Following standard metrics in the field \citep[e.g.,][]{Dey2021Capsule}, we evaluate our model on both point-like metrics and evaluate the probability estimates given by our model via a visual calibration metric. For the point based metrics, we define the following using the set of scaled residuals $r = {\frac{z_i - \hat{z}_i}{(1+z_i)}}$, and allow $r_i$ to be the $i$-th element of that set. In the following let $\text{med}(\cdot)$ represent the median over a set. We define a robust measure of spread, the scaled median absolute deviation (\textbf{MAD}) $= 1.4826 \times \text{med}(|r_i - \text{med}(r_i)|)$, the bias of the residuals (\textbf{BIAS}) $= \frac{1}{|r|} \times \sum_{i=1}^{|r|} r_i$, and the percentage of catastrophic outliers ($\mathbf{\eta}$) defined to be the percentage of scaled residuals with values greater than 0.05. We evaluate probabilistic model calibration using the field-standard probability integral transform (\textbf{PIT}) \citep{PITOGDawid1984}, described and reported in appendix~\ref{appendix:PIT}. We evaluate our model on the held-out test dataset only after making architecture and hyperparameter choices via performance on the validation set. 

%We benchmark our model using our test dataset and using inputs with targets from the Sloan Digital Sky Survey main galactic sample \citep{SDSS_MGS} (SDSS:petro\_r < 17.77).

%The metric of choice for model calibration in the field is the probability integral transform (\textbf{PIT}) (Schmidt or Malz). The PIT is a histogram of the occurrences of true redshift in the set of CDFs measured from density estimates $z_i$, or $\text{CDF}(z_i,y_i) = \int_0^{y_i} z_i \mathrm{d}z$. As PIT is a visual metric, we report the PIT in appendix~\ref{appendix:PIT}.

\textbf{Interpretability:}
\label{sec:interpretability_methods} \label{sec:shapley}
 %Background on Shapley Values
Shapley values quantify the value of each input feature in a cooperative game, commonly framed as the learning target of a neural network  \citep{ShapleyOG, ShapleyRepopularizedforNN}. A formal definition is provided in appendix~\ref{appendix:shapley}, but here we briefly build some intuition. For each datapoint, we modify each input bands and measure the resultant change in the predicted redshift. To modify an input band, we replace the image with Gaussian noise, where the mean and variance are sampled from robust measures of the background's moments in each image. This modified input physically represents a ``non-detection'' for that band, since galaxies appear as bright sources on a noisy background. We then sample over every combination of non-detections in each band. The average effect of all samples with the feature included is compared to samples with the feature excluded to model its utility to point estimation. For each band, a larger Shapley value means a larger change in the predicted output. Positive values reflect predictions are shifted to higher redshifts, while negative values reflect a shift to lower redshifts when the band is present. The multimodal Shapley value (MM-SHAP) \citep{WhichInputMultiModal}, ranks the relative importance of each input band. Both the Shapley and $\text{MM-SHAP}$ values are calculated for each datapoint; we monitor the central value of the distribution in bins of true redshift. This allows us to connect the most useful bands to underlying spectral features.

%For a single datapoint, the Shapley value of each filter is $f_j$, and the MM-SHAP value of that filter is then normalized relative to the other filters: $\text{MM-SHAP}(f_j) = \frac{|\phi_j|}{\sum_j^F |\phi_j|}$. Note, because Shapley and $\text{MM-SHAP}(f_j)$ are calculated for each $x_i$, we monitor the distribution of $\text{MM-SHAP}(f_j)$ over bins in true redshift. This allows us to connect the most useful bands to underlying spectral features, which we can use to evaluate the utility of the interpretability metrics.

\section{Results and Discussion}
\textbf{Performance:}
Model performance is shown in Table~\ref{tab:main_reults}, including scaled median absolute deviation (MAD), the bias of the residuals, and the percentage of catastrophic outliers ($\eta$). The \textit{Mantis Shrimp} multi-instrument CNN model is compared against several literature benchmarks, including an optical-only catalog-based model \citep{Beck2016LLNSDSS}, optical-only CNN models \citep{Pasquet2019, Dey2021Capsule, Hayat_2021_selfsupervised}, and a multi-instrument catalog-based model \citep{WISE-PS1-STRMBeck22}. Performance on our own held-out test split is shown in the top row. The comparison with literature benchmark is performed on matched evaluation sets: SDSS MGS for the optical-only models and PS1 $\times$ UnWISE for the multi-instrument model. Additional details regarding the creation of these evaluation sets is available in appendix~\ref{appendix:comparison_details}.

The performance of the \textit{Mantis Shrimp} multi-instrument CNN model does not yet match the performance of the best-performing optical-only multi-instrument models. However, we note that the model presented here is still preliminary: it has only been trained with $\sim10\%$ of the total dataset and we have not yet done rigorous hyperparameter tuning. We remain optimistic that we can continue to improve performance. When comparing to optical-only models on SDSS MGS: the \textit{Mantis Shrimp} model outperforms the catalog-based model in MAD, and has less residual bias than the catalog-based model and two of the three CNNs. When comparing to the multi-instrument model, \textit{Mantis Shrimp} model shows fewer catastrophic outliers than the catalog-based model.
%Earlier works investigating CNNs for photo-z including  train and benchmark using this population. We also compare to the population of galaxies with catalog photometry available in PS1 $\times$ UnWISE, (the WISE-PS1-STRM catalog) (B22), as it represents the most similar work through incorporation of both IR and optical features in the same region of sky. 

\begin{table}[!ht]
\caption{\small\textbf{Model Performance:} We compare \textit{Mantis Shrimp} model (this work) to prior works. The asterisk (*) marks metrics as reported by the original authors, rather than computed directly.}
  \centering
 \label{tab:main_reults}
  \begin{tabular}{llcccccc}
    \toprule
    Evaluation Data & Model & MAD & Bias & $\eta$ (\%) \\
    \hline
    \multirow{3}{*}{Test Set}  & Multi-instrument CNN (This Work) & \textbf{0.0244} & 1e-2 & \textbf{13.84}  \\
    & Optical and IR CNN (This Work) & 0.0250 & \textbf{9e-3} & 14.24  \\
    & Optical only CNN (This Work) & 0.0342 & 1e-2 & 20.08  \\
    \hline
    \multirow{5}{*}{SDSS MGS} & Multi-instrument CNN (This Work) & 0.0114 & \textbf{1e-5} & 1.05  \\
      & Optical-only \citep{Beck2016LLNSDSS} & 0.0140 & 9e-4 & 1.60   \\
     & Optical-only CNN \citep{Pasquet2019}* & 0.0091 & 1e-4 & 0.31    \\
    & Optical-only CNN \citep{Hayat_2021_selfsupervised}* & \textbf{0.0083} & 1e-4 & 0.21 \\
      & Optical-only CNN \citep{Dey2021Capsule}* & 0.0089 & 7e-5 & \textbf{0.19}   \\
     \hline
     \multirow{2}{*}{PS1 $\times$ WISE}  & Multi-instrument CNN (This Work) & 0.0210 & 3e-3 & \textbf{8.08}  \\
      & Multi-instrument \citep{WISE-PS1-STRMBeck22} & \textbf{0.0185} & \textbf{2e-3} & 8.28   \\
     \hline
    \bottomrule
  \end{tabular}
\end{table}

\begin{figure}[!ht]
    \centering
    \includegraphics[scale=0.38]{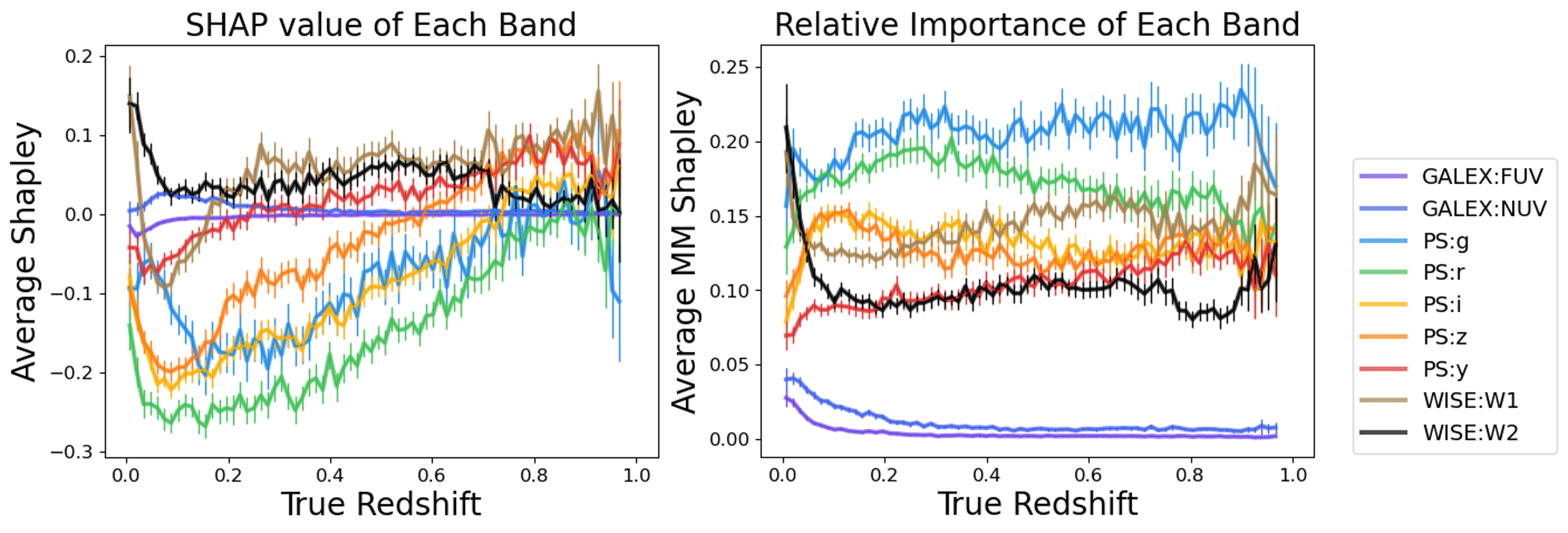} 
    \caption{\small\textbf{Shapley (left) and MM-SHAP (right) values for each band, averaged across the samples in each redshift bin.} We find that the model prioritizes information from the three instruments differently, and that prioritization changes as a function of redshift. This is inline with our understanding of the underlying physics, see appendix~\ref{appendix:physics} for a review.}
    \label{fig:interp}
\end{figure}

\textbf{Interpretability:}
We compute both the original Shapley and the normalized MM-SHAP scores for each channel as outlined in section \ref{sec:shapley}, and visualize the results in figure~\ref{fig:interp}. We find that the model prioritizes information from each instrument differently, and that prioritization changes as a function of redshift. This is inline with our understanding of the underlying physics, as key spectral features move to longer and redder wavelengths as redshift increases. See appendix \ref{appendix:physics} for a review. The movement of a spectral features through different observed bands has long been attributed to as the reason for success or failure in photo-z estimation, especially in template fitting approaches,\citep{4000Abreak_affectsPZ, ANNz2_and4000A_break_Affects_PZ}, and provides useful context for interpreting widely used ``interpretability'' metrics. 
%\emph{Left (original Shapley values)}: at redshifts below $\sim0.6$, non-detections in the optical bands (g,r,i,z,y) cause the model to under-predict redshifts, while non-detections in the IR bands (W1,W2) cause the model to over-predict redshifts. At high redshifts ($\sim1.0$), non-detections in any band cause the model to over-predict redshift. Non-detections in the UV bands have little effect on the predicted redshift. \emph{Right (MM-SHAP values)}: out of the 9 input bands, both of the UV bands are considered the least informative at all redshifts. The optical bands (in particular the g,r,i bands) are the most important bands below a redshift of $\sim0.5$, after which, the IR bands are the most important bands.

%There is value in reporting and visualizing both metrics, as the Shapley value sign tells us how adding the channel influences the point estimates (increasing or decreasing them), while the MM-SHAP scores are normalized to more easily compare the relative importance of each band at each redshift. We visualize the results in figure~\ref{fig:interp}.

%For each band, we plot the mean Shapley or MM-SHAP value across samples in each redshift band. The Shapley values communicate the effect on the point estimate of redshift if that filter is excluded from the model. The MM-SHAP communicates the relative importance of each filter compared to one-another at each redshift interval.
%We actually can't say anything about surface brightness since that is a derived feature of the entire channel. We simply get importances. 
%\section{Discussion}
%Conclusion. First, what can we say about the evidence we have gathered so far re. interpretability.

The MM-SHAP indicates that the model relies heavily on the optical g and r bands for accurate redshift estimation throughout the interval. At nearby redshifts, there is a strong drop in the relative importance of the IR bands with a strong initial positive Shapley value. This reflects that the typical galaxy SED curve become brighter in the IR with increased redshift, see appendix ~\ref{appendix:physics} for a visualization.  

%We suggest that this reflects the shift of an important and bright spectral feature from g-band to r-band at $z=0.4$. Moreover, it also helps explain the sharp decline in the g-band Shapley values. Galaxies at redshifts above 0.4 have a dim g-band and a bright r-band; thus, non-detections in the g-band have less of an impact at redshifts above $0.4$, which is also reflected in the Shapley values. The Shapley metrics also appropriately reflect the utility of including IR bands, particularly for objects at high redshift. In both Shapley metrics, the IR bands are relatively important across all bins of true redshift, with an increased importance at high redshift.
Overall, the MM-SHAP values confirm that both the IR and Optical filters are being utilized by our network and trade-off in importance in different redshift intervals. In contrast, the Shapley values convey that the UV bands are relatively un-important at all redshifts. While we expected that the UV bands would be incredibly useful from a color perspective, the relative depth of Galex combined with the fact that many galaxies are inherently UV dim from understood galaxy population statistics (fewer young stars, which emit copious amounts of UV light) suggests otherwise. To evaluate this hypothesis, we re-trained the CNN with UV bands removed, and UV+IR bands removed, and report the results in Table~\ref{tab:main_reults}, and there do verify that 1) UV bands can be removed without affecting our performance, and 2) in contrast that removing the IR bands \textit{is detrimental} to performance.  

%In contrast to the UV bands, we see  The most notable feature of figure\ref{fig:interp} (a) are that the WISE channels always have on average positive Shapley value across bins of true redshift, while the PanSTARRS survey channels are typically negative, especially at low redshift. 
%Then add a small paragraph about the meaning of our results on performance.
%The performance of our results against a few different benchmark datasets are so far less performant against the most comparable PS1 $\times$ WISE catalogue of galaxies from available from B22. Considering that B22 estimates are already available as a public catalogue and would have much lower computational cost compared to our ResNet model, we likely must substantially exceed performance to motivate the use of our model. Reflecting on our own progress, we are nonetheless optimistic due to the fact we are only using 7\% of the total available dataset in this report and have not performed any automated hyperparamter tuning. 

%As this report details preliminary work, we end by soliciting feedback and share our intention for future work. 
This work details preliminary efforts to ascribe photo-z CNN model behavior to physics underlying galaxy SEDs. We end by soliciting feedback and sharing our intention for future work. We must include the remaining 93\% of available data for our training set, perform automated hyperparameter tuning runs, and will look to evaluate different approaches for fusing the imagery across instruments. Future work should also improve upon our method for defining a baseline image. A fundamental limitation of our work is Shapley values do not distinguish between the effects of SED movement (i.e., color-shifts) from the surface-brightness and distance relationship. Works that seek to extract signatures in CNN models from specific known physics, like the Balmer break, will need to overcome this fact.

%Finally, a quirk of our method is that each unique view of a patch of sky is mapped to a redshift estimate. Since galaxies are typically extended across multiple pixels, re-centering the images within a small size along the semi-major an semi-minor axes of galaxies might be a sensible and meaningful augmentation, perhaps even necessary to constrain our model's variance over such shifts from noisy astrometry or a user's mis-specification.

%Limiations
%A fundamental limitation of the work is whether the population of spectroscopically identified galaxies are reflective of the population of all galaxies. Due to the targeting bias of individual surveys and observational bias from Malmquist effect, it is known that test datasets randomly sampled from training sets have distribution shifts compared to all observable galaxies. Caution must be used specifically when applying our model to data outside the targeting criteria of the surveys used. A second limitation of our interpretability approach is the ambiguity of the definition of the baseline image. The most correct baseline image would be to remove the light contributed to each pixel from only the target galaxy, leaving much of the original image unchanged. Future work could look to isolate the light from target galaxies to create a better baseline image.

%Future Work-- we can limit these.

\bibliographystyle{./neurips.bst}
\bibliography{bibliography}

\appendix

\section{Acknowledgements}
%People Acknoledgements. 
%Comment out for anonymous review
This work would not have been possible without correspondence from Dustin Lang (Perimeter Institute for Theoretical Physics) and Aaron Meisner (NSF NOIRLab) on help with querying for UnWISE and GALEX images from the Legacy Survey. Additionally, thanks to Rick White and Travis Berger, (Space Telescope Science Institute) for help querying PanSTARRS image stacks. A. Engel and N. Byler were partially supported by an interagency agreement (IAA) between NASA and the DOE in liu of grant awarded through the NASA ROSES D.2 Astrophysics Data Analysis grant\# 80NSSC23K0474, ``Multi-Survey Photometric Redshifts with Probabilistic Output for Galaxies with 0.0 < Z < 0.6.'' PNNL is a multi-program national laboratory operated for the U.S. Department of Energy (DOE) by Battelle Memorial Institute under Contract No. DE-AC05-76RL0-1830.

%Photometric
The Pan-STARRS1 Surveys (PS1) and the PS1 public science archive have been made possible through contributions by the Institute for Astronomy; the University of Hawaii; the Pan-STARRS Project Office; the Max-Planck Society and its participating institutes, the Max Planck Institute for Astronomy, Heidelberg and the Max Planck Institute for Extraterrestrial Physics, Garching; The Johns Hopkins University, Durham University; the University of Edinburgh; the Queen’s University Belfast; the Harvard-Smithsonian Center for Astrophysics; the Las Cumbres Observatory Global Telescope Network Incorporated; the National Central University of Taiwan; the Space Telescope Science Institute; the National Aeronautics and Space Administration under grant no. NNX08AR22G issued through the Planetary Science Division of the NASA Science Mission Directorate; the National Science Foundation grant no. AST-1238877; the University of Maryland; Eotvos Lorand University (ELTE); the Los Alamos National Laboratory; and the Gordon and Betty Moore Foundation.

%UnWise
The Legacy Surveys consist of three individual and complementary projects: the Dark Energy Camera Legacy Survey (DECaLS; Proposal ID \#2014B-0404; PIs: David Schlegel and Arjun Dey), the Beijing-Arizona Sky Survey (BASS; NOAO Prop. ID \#2015A-0801; PIs: Zhou Xu and Xiaohui Fan), and the Mayall z-band Legacy Survey (MzLS; Prop. ID \#2016A-0453; PI: Arjun Dey). DECaLS, BASS and MzLS together include data obtained, respectively, at the Blanco telescope, Cerro Tololo Inter-American Observatory, NSF’s NOIRLab; the Bok telescope, Steward Observatory, University of Arizona; and the Mayall telescope, Kitt Peak National Observatory, NOIRLab. Pipeline processing and analyses of the data were supported by NOIRLab and the Lawrence Berkeley National Laboratory (LBNL). The Legacy Surveys project is honored to be permitted to conduct astronomical research on Iolkam Du’ag (Kitt Peak), a mountain with particular significance to the Tohono O’odham Nation.

NOIRLab is operated by the Association of Universities for Research in Astronomy (AURA) under a cooperative agreement with the National Science Foundation. LBNL is managed by the Regents of the University of California under contract to the U.S. Department of Energy.

The Legacy Survey team makes use of data products from the Near-Earth Object Wide-field Infrared Survey Explorer (NEOWISE), which is a project of the Jet Propulsion Laboratory/California Institute of Technology. NEOWISE is funded by the National Aeronautics and Space Administration.

%GALEX?
We gratefully acknowledge NASA’s support for construction, operation, and science analysis for the GALEX mission, developed in cooperation with the Centre National d’Etudes Spatiales of France and the Korean Ministry of Science and Technology. The grating, window, and aspheric corrector were supplied by France. We acknowledge the dedicated team of engineers, technicians, and administrative staff from JPL/Caltech, Orbital, University of California, Berkeley, Laboratoire d’Astrophysique de Marseille, and the other institutions who made this mission possible.

%Spectroscopy - SDSS and DESI
Funding for the DEEP2 Galaxy Redshift Survey has been provided by NSF grants AST-95-09298, AST-0071048, AST-0507428, and AST-0507483, as well as NASA LTSA grant NNG04GC89G. This research uses data from the VIMOS VLT Deep Survey, obtained from the VVDS database operated by Cesam, Laboratoire d’Astrophysique de Marseille, France. This paper uses data from the VIMOS Public Extragalactic Redshift Survey (VIPERS). VIPERS has been performed using the ESO Very Large Telescope, under the ‘Large Programme’ 182.A-0886. The participating institutions and funding agencies are listed at http://vipers.inaf.it. The WiggleZ survey acknowledges financial support from The Australian Research Council (grants DP0772084, LX0881951, and DP1093738 directly for the WiggleZ project, and grant LE0668442 for programming support), Swinburne University of Technology, The University of Queensland, the Anglo-Australian Observatory, and The Gregg Thompson Dark Energy Travel Fund at UQ. GAMA is a joint European-Australasian project based around a spectroscopic campaign using the Anglo-Australian Telescope. The GAMA input catalogue is based on data taken from the Sloan Digital Sky Survey and the UKIRT Infrared Deep Sky Survey. Complementary imaging of the GAMA regions is being obtained by a number of independent survey programmes including GALEX MIS, VST KiDS, VISTA VIKING, WISE, Herschel-ATLAS, GMRT and ASKAP providing UV to radio coverage. GAMA is funded by the STFC (UK), the ARC (Australia), the AAO, and the participating institutions. The GAMA website is http://www.gama-survey.org/. This paper made use data from the Final Release of 6dFGS. the 6dFGS website is http://www-wfau.roe.ac.uk/6dFGS/.

%SDSS
Funding for the Sloan Digital Sky Survey IV has been provided by the Alfred P. Sloan Foundation, the U.S. Department of Energy Office of Science, and the Participating Institutions. SDSS-IV acknowledges support and resources from the Center for High-Performance Computing at the University of Utah. The SDSS website is www.sdss.org.

SDSS-IV is managed by the Astrophysical Research Consortium for the Participating Institutions of the SDSS Collaboration including the Brazilian Participation Group, the Carnegie Institution for Science, Carnegie Mellon University, the Chilean Participation Group, the French Participation Group, Harvard-Smithsonian Center for Astrophysics, Instituto de Astrofísica de Canarias, The Johns Hopkins University, Kavli Institute for the Physics and Mathematics of the Universe (IPMU) / University of Tokyo, the Korean Participation Group, Lawrence Berkeley National Laboratory, Leibniz Institut für Astrophysik Potsdam (AIP), Max-Planck-Institut für Astronomie (MPIA Heidelberg), Max-Planck-Institut für Astrophysik (MPA Garching), Max-Planck-Institut für Extraterrestrische Physik (MPE), National Astronomical Observatories of China, New Mexico State University, New York University, University of Notre Dame, Observatário Nacional / MCTI, The Ohio State University, Pennsylvania State University, Shanghai Astronomical Observatory, United Kingdom Participation Group, Universidad Nacional Autónoma de México, University of Arizona, University of Colorado Boulder, University of Oxford, University of Portsmouth, University of Utah, University of Virginia, University of Washington, University of Wisconsin, Vanderbilt University, and Yale University.

%DESI
This research used data obtained with the Dark Energy Spectroscopic Instrument (DESI). DESI construction and operations is managed by the Lawrence Berkeley National Laboratory. This material is based upon work supported by the U.S. Department of Energy, Office of Science, Office of High-Energy Physics, under Contract No. DE–AC02–05CH11231, and by the National Energy Research Scientific Computing Center, a DOE Office of Science User Facility under the same contract. Additional support for DESI was provided by the U.S. National Science Foundation (NSF), Division of Astronomical Sciences under Contract No. AST-0950945 to the NSF’s National Optical-Infrared Astronomy Research Laboratory; the Science and Technology Facilities Council of the United Kingdom; the Gordon and Betty Moore Foundation; the Heising-Simons Foundation; the French Alternative Energies and Atomic Energy Commission (CEA); the National Council of Science and Technology of Mexico (CONACYT); the Ministry of Science and Innovation of Spain (MICINN), and by the DESI Member Institutions: www.desi.lbl.gov/collaborating-institutions. The DESI collaboration is honored to be permitted to conduct scientific research on Iolkam Du’ag (Kitt Peak), a mountain with particular significance to the Tohono O’odham Nation. Any opinions, findings, and conclusions or recommendations expressed in this material are those of the author(s) and do not necessarily reflect the views of the U.S. National Science Foundation, the U.S. Department of Energy, or any of the listed funding agencies.
 
%PLANCK for dust map.
The Planck dust map is based on observations obtained with Planck (http://www.esa.int/Planck), an ESA science mission with instruments and contributions directly funded by ESA Member States, NASA, and Canada.

\section{Related Work}
\label{appendix:related_work}

\textbf{Photometric Redshift Estimation.}
\label{appendix:related_work_PZ}

%What are the previous works
Previous works on photo-zs have investigated the use of various ML algorithms \citep{ANNz2_and4000A_break_Affects_PZ,ANNzCollister2003,SDSS_PhotoZ_GaussianProcesses,Beck2016LLNSDSS}, but of particular interest to this article are works that utilized computer vision models (CNNs) to estimate the redshift directly from the original science images, rather than from derived features available in catalogues \citep{Pasquet2019,Dey2021Capsule,Hayat_2021_selfsupervised,CNN_multichannel_PZ_DIsanto2018,FirstCNNHoyle2015,CNNSDSSLargeHenghes2021}. In these works, computer vision networks show replicable improved performance across the range of community accepted metrics, summarized in section~\ref{sec:metrics}. In particular, P19 is an influential example. Evaluation of photo-z computer vision models have been limited to the SDSS Main Galactic Sample. While this is an important benchmark dataset, we ask whether computer vision networks remain useful in a larger volume of input-redshift space. 

A second line of work that serves as background for our model is the combination of multiple surveys spanning different wavelengths of the electromagnetic spectrum \cite{WISE-PS1-STRMBeck22} (B23). They showed that the combination of UnWISE in the near infrared (NIR) with the PanSTARRS survey in the visual wavelengths create an improvement over their previous work using PanSTARRS alone \citep{Beck21PS1STRM}. We extend this work by incorporating the additional GALEX survey, which includes information in the ultra violet (UV). In addition, an added benefit of our approach is that we do not complicate data collection with catalogue joins. A fundamental limitation to catalogue joins is that a source must be detected (typically defined as a 5$\sigma$ pixel value) in a survey band to register as an object. B21 goes to great lengths describing their careful catalogue joins. We can circumvent this issue by simply querying an image cutout service at the desired coordinates.

\textbf{Interpretability In Photometric Redshift Estimation}
\label{appendix:related_work_Interp}
% Many works have sought to explain neural networks to users, with different approaches generally seen attributing behavior to different phenomena. Among these are gradient of activation based approaches including CAM \citep{CAM}, GradCAM \citep{GradCAM}, LIME \citep{LIME}. Popular work from \citet{OlahInterpretability} take a different approach where an attribution to individual/combinations of convolutional filters are visualized by perturbing inputs to maximize the activation. Gradients of the output with respect to the input image itself have been used to localize importance to the input image \citep{VanillaGradients2013}, and more recently incorporate a baseline image representing no-feature contribution \citep{IntegratedGradients2017}. More recently, kernel methods have been used to attribute decision to specific datapoints \citep{anonymous2024faithful}. Finally, the Shapley value (which this work employs) has been re-popularized for neural networks \citep{ShapleyRepopularizedforNN}. We provide a more thorough introduction to Shapley value in \ref{appendix:shapley}.   
Previous works have sought to explain the performance of empirical photo-z algorithms through investigation of feature importance \citep{FeatureImportanceDIsanto}, and through the use of capsule networks \citep{CapsuleNetworkSabour2017} to probe generative model output at varying embedded feature scale \citep{Dey2021Capsule}. The latter concluded that the most important features to a computer vision network trained on the SDSS MGS were a complex feature spanning morphology (elliptical vs spiral) and orientation, while the second most important feature was heavily correlated to surface brightness. We will differ from these past works by specifically targeting the relationship between redshift point-prediction and individual astronomical filters.

\textbf{Multimodal or Fusion Networks.}
\label{appendix:related_work_Fusion}
This paper studies the combination of multiple broad-band filters for the task of photometric redshift estimation. Each filter represents a different slice of the electromagnetic spectrum (see figure~\ref{fig:physics2}), with different characteristic phenomena recorded for any redshift. Additionally, each survey has its own instrumentation with different sensors. For this reason, we frame our study as a data fusion experiment where multiple modalities are combined together \citep{MultiModalityReview}. Relevant topics in data fusion include what architecture performs best given these multiple input types (e.g., late vs early fusion \citep{Early_vs_Late_CNN}), how to ensure each data modality is being utilized rather than the model collapsing out any modality \citep{UniModal_Collapse}, and which modality is most important \citep{Godfrey23MultiSpectral}. It is this literature that motivated us to ask that main research question of this article: ``which filter are most important in each redshift bin?''

\section{Full Background on Galaxy Radiation}
\label{appendix:physics}

We can model the total luminosity observed $\Phi$ from an object in any particular filter with transmission curve $R(\lambda)$ and object spectral energy density (SED) $F_\lambda$ as \citep{ABMagnitudeSystem}:

\begin{equation*}
\Phi = \frac{\int_{\lambda_a}^{\lambda_b} F(\lambda) R(\lambda) \lambda \mathrm{d}\lambda}{\int_{\lambda_a}^{\lambda_b}  c \frac{1}{\lambda} R(\lambda) \mathrm{d}\lambda}
\end{equation*}

\begin{figure}[!ht]
    \centering
    \includegraphics[scale=0.39]{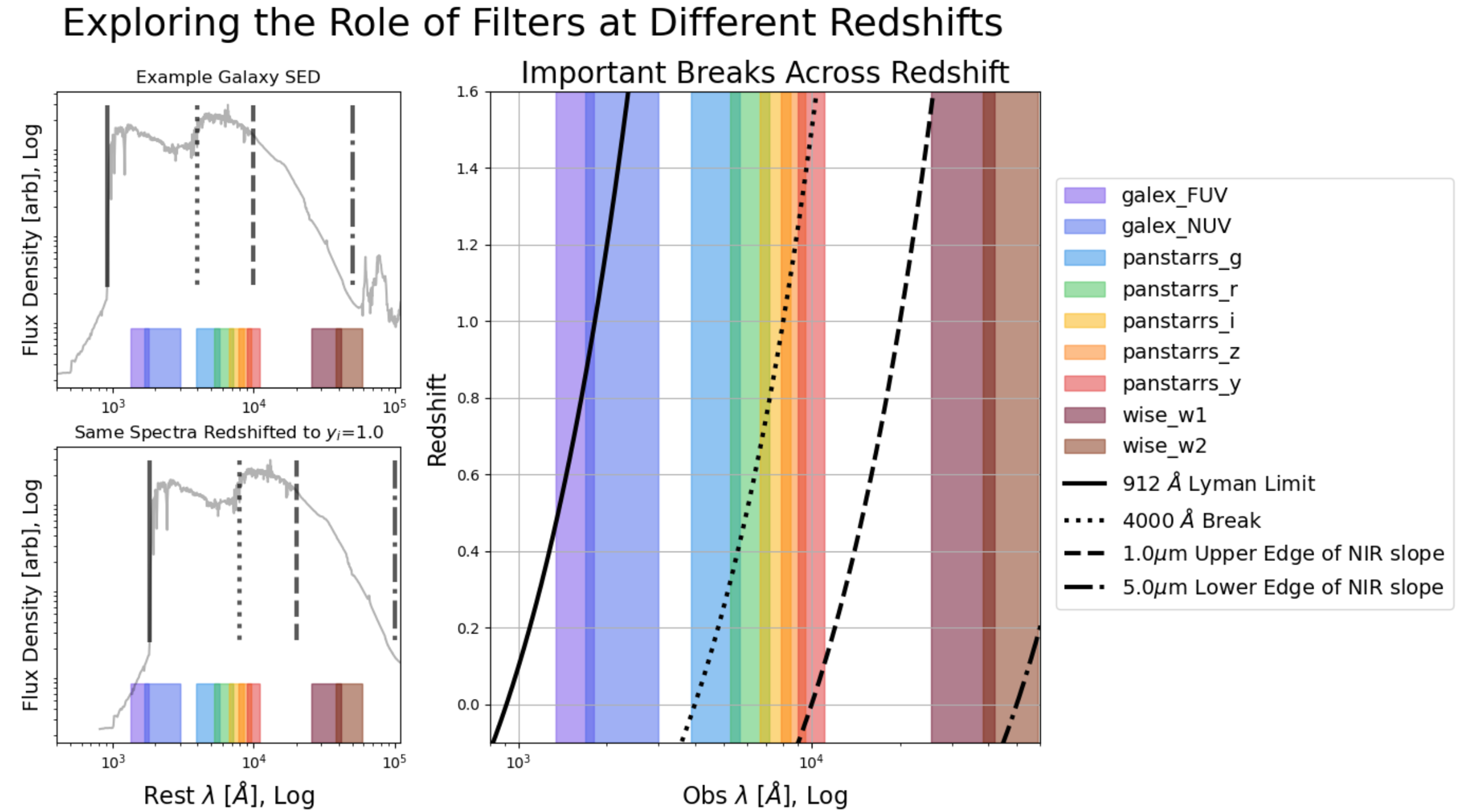}
    \caption{\textbf{Intuition about which channels should be important when}.  \textbf{Top Left.} A representative galaxy SED (taken from the atlas in \citet{Brown14_galTemplates}) is plot at that galaxy's rest-frame, with the photometric filters used in this work overlaid. Additionally, important breaks in the galaxy SED are identified with vertical black lines. \textbf{Lower left.} We plot the same SED except redshifted to a value of 1.0 to visually represent how light incoming from a galaxy appears in our observation frame. The important breaks/phenomena have shifted with the spectra and now coincide in different parts of filter-space. \textbf{Right.} Finally, we plot each of the SED Break characteristic wavelength as a function of redshift. These curves pick out at which filters we expect the be important at each redshift. A description of these breaks are given in appendix~\ref{appendix:physics}. Finally, note the filter regions are shown as uniform boxes for visual ease-- the actual filters have unique, bell shaped transmission curves.}
    \label{fig:physics2}
\end{figure}

The observed SED shifts with redshift as $\lambda_{obs} = \lambda_{emit}(1+z)$, but the filter transmission curves stay fixed in our observation frame. This means that phenomena in galaxy SED move successively through series of filters with redshift. This process has long been attributed to as an explanation for when/where template fitting photo-z techniques succeed or fail \citep{4000Abreak_affectsPZ}. Since this intuition ties the performance of the model to the object's redshift and specific filters, we ask whether the same could be applied to understand neural network behavior. We explore this in figure~\ref{fig:physics2}. 

In figure~\ref{fig:physics2} we identify how the Lyman limit (912\AA), a ``break'' or sudden change in the SED and caused by the ionization energy of the neutral hydrogen atom, serves as the upward slope of a typical galaxy SED. The source of that UV light is from an abundance of bright and short-lived stars, which occur only around star forming regions. We also identify the 4000\AA~break, which is particularly prominent in galaxies without many young blue stars. It is caused by the ionized metal in the atmospheres of stars that absorb radiation above 4000\AA~ \citep{4000Break_explanation}. 

Our final region is the long gently sloping tail from 1$\mu$m-5$\mu$m. It is known that this gentle slope is a common feature in the (rest-frame) NIR to all galaxy types \citep{Mannucci2001NIRsame}. This is in contrast to the Lyman Limit and 4000\AA~break, whose prominence in galaxies are sensitive to the star formation history of the galaxy, which is strongly correlated to galaxy type. Our intuition is that: as an SED becomes redshifted, available light leaves the UV/Optical bands, but the luminosity of the NIR bands increases.

\section{Definition of Neural Networks and Photometric Redshift Estimation Task}
\label{appendix:neural_networks}

Formally, let $D$ be a labeled dataset of length $N$ composed of individual data-label pairs drawn from the population of spectroscopically identified galaxies, $(x_i, z_i) \in (\mathcal{X},\mathcal{z}) \in (R^{F \times H \times W}, )$ such that $D=\{(x_1, z_1),(x_2, z_2),\ldots,(x_N, z_N)\}$. We will find it useful to define $\mathfrak{Z}$ as the set of vector with number of elements C where each element represents a small bandwidth of redshift value, linearly spanning the space of $[\min_i(z_i),\max_i(z_i)]$. We will further notate $\mathfrak{Z}^j$ as the j-th element of the vector $\mathfrak{Z}$. Our neural network function $F$ is a parameterized and learnable mapping from galaxy photometry to a redshift latent vector, $F: \mathcal{X} \to \mathcal{Y} \in \Delta^C$, in the space of unit simplex vectors with dimension C (achieved through the SoftMax activation function). Let $\theta$ be the vector of learnable parameters. We will adopt the methodology of P19, and assume $y^j_i \approx P(\hat{z}_i \in \mathfrak{Z}^j + \delta z |x_i,)$, which we will evaluate empirically (appendix~\ref{appendix:PIT}). The weights $\theta$ are learned via gradient descent to minimize the cross entropy loss evaluated between $y_i$ and $1^c$, where $1^c$ is the one-hot encoded vector of integer $c$ representing the correct bin of $\mathfrak{Z}$ that the true redshift $z_i$ falls into. Using this interpretation of $y_i$, we can measure a point estimate through measuring the expectation value $\hat{z}_i = \mathbb{E}(y_i)$. 

\section{Spectroscopic Data Collection}
\label{appendix:spectroscopic}

\begin{table}[!ht]
\caption{\small\textbf{Spectroscopic Band Table} $N_\text{avail}$ is the total amount of data available after cuts. $N_{\text{samples}}$ is the number of accepted samples in the training dataset. Notes are the quality cuts applied to each survey.}
  \centering
 \label{tab:data_table}
  \resizebox{\textwidth}{!}{\begin{tabular}{lllll}
    \toprule
    Source & Ref & $N_{\text{avail}}$ & $N_{\text{samples}}$  & Notes\\
 \hline
     SDSS & \citep{SDSS_four} & 5.1e6 & 2.6e6 & ZWARNING=0,  Class=Galaxy\\
     DESI & \citep{DESI_EDAspectroscopy} & 2.4e6 & 1.1e6 & ZWARN=0, SPECTYPE=GALAXY\\
     DEEP2 & \citep{DEEP2} & 5.3e4 & 3.4e4  & q\_z>2,Cl=G,\\
     GAMA & \citep{GAMA_DR2} & 3.4e5 & 2.0e5 & NQ>=3\\
     VVDS & \citep{VVDS_final_release} & 40,944 & 10,296  & zflags=3 or zflags=4\\
     VIPERS & \citep{VIPERS_final} & 6.0e4 & 4.9e4 & classflag=0 or 1, zflg<5 \& zflg>=2\\
     6dF & \citep{6dFGS_final} & 1.2e5  & 5.6e4 & QUALITY=3 or QUALITY=4\\
     WiggleZ & \citep{WiggleZ_final} & 2.2e5 & 1.4e5 & Q>=4\\
    \bottomrule
  \end{tabular}}
\end{table}

Spectroscopic samples were compiled from available surveys and combined together, including the Sloan Digital Sky Survey (SDSS) \cite{SDSS_four} at data release 17. We make effort to include \textbf{all} available spectroscopy satisfying our quality cuts, including that of the Main Galaxy Survey (MGS) \cite{SDSS_MGS}, the Baryonic Oscillation Spectroscopic Survey (BOSS) \cite{BOSS_target_guidelines}, and the extended Baryonic Oscillation Spectroscopic Survey (eBOSS) \citep{EBOSS}. We include the Dark Energy Spectroscopic Instrument survey DESI \citep{DESI_EDAspectroscopy} where at time of writing, only the early data release is available. SDSS and DESI make up the vast majority of our sample. We also include measurements from DEEP2 \citep{DEEP2}, GAMA \citep{GAMA_DR2}, VVDS \citep{VVDS_final_release}, VIPERS \citep{VIPERS_final}, 6dF \citep{6dFGS_final}, and the WiggleZ \citep{WiggleZ_final} surveys. Information on the quality cuts used for each survey are provided in table \ref{tab:data_table}, including the final total amount of ground truth labels taken from each survey. Our quality cuts and compilation of survey data closely match that of \citep{Beck21PS1STRM}; our main advantage is the inclusion of the sizable DESI EDA spectroscopic sample, adding some 1Million additional targets for training.

The targeting guidelines of each of the spectroscopic surveys essentially create a biased and incomplete sample of the input feature space with respect to the entire population of observable galaxies \cite{NewmanSpectroscopicBias}. We do not consider here how to correct for or flag examples that fall outside the support of the training dataset; however, In the next few paragraphs we provide additional details of the query and processing of the spectroscopic target data for each source, including any cuts we make. We also document the populations studied in each survey.

\begin{figure}[!ht]
    \centering
    \includegraphics[scale=0.5]{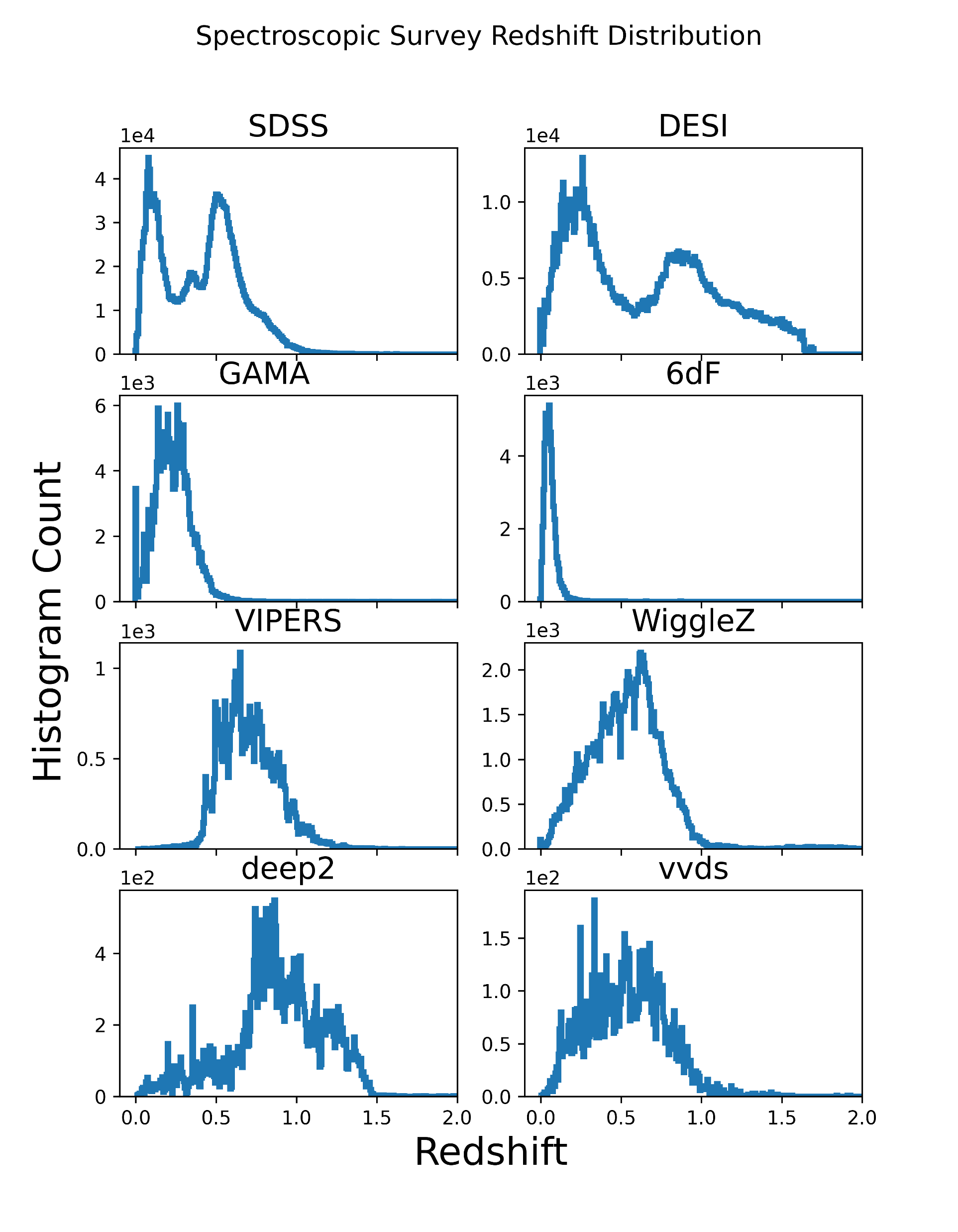}
    \caption{\small\textbf{Distribution of targets in redshift visualized by individual survey}}
    \label{fig:spectroscopicdist}
\end{figure}

\begin{figure}[!ht]
    \centering
    \includegraphics[scale=0.5]{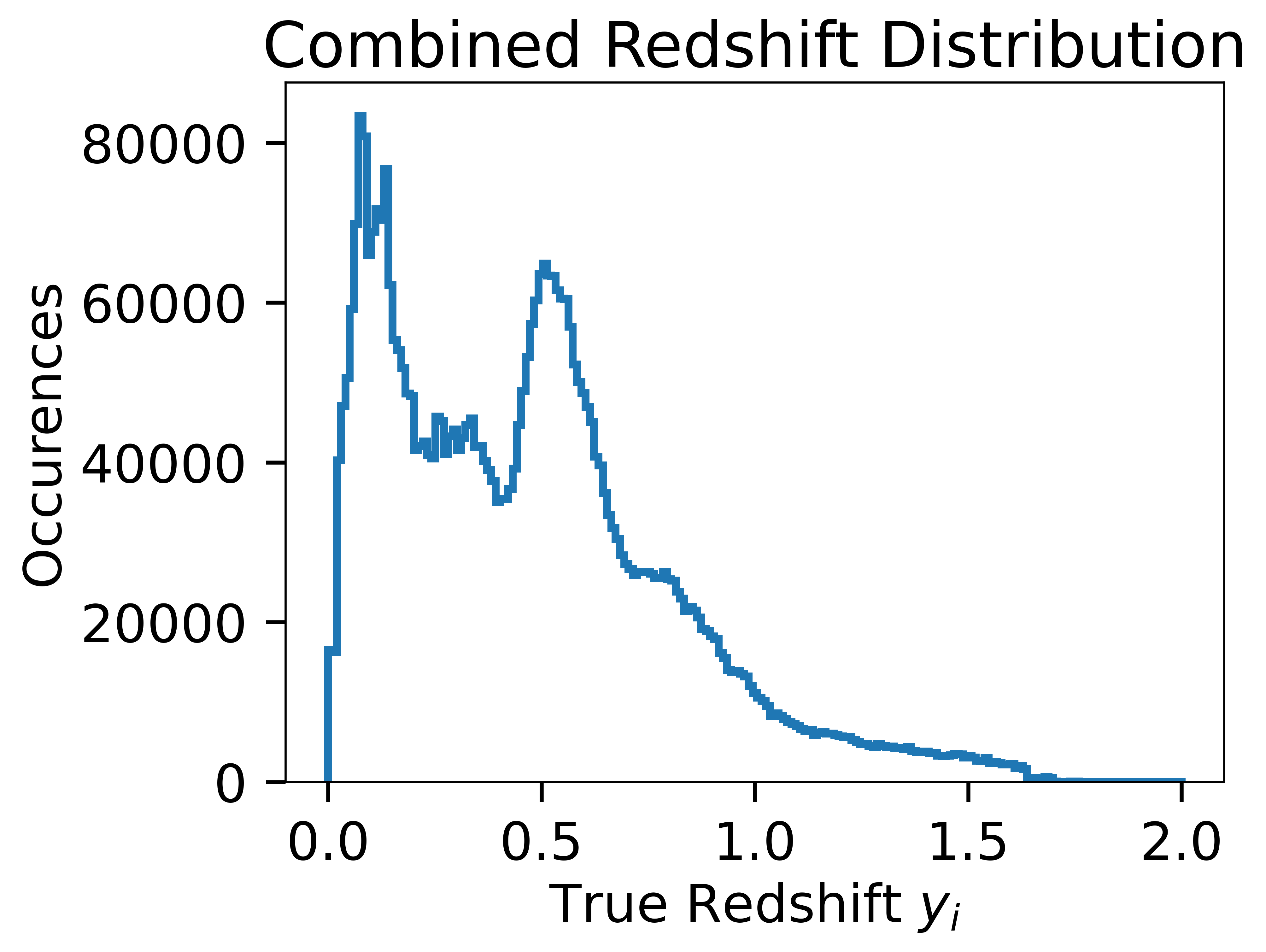}
    \caption{\small\textbf{Distribution of targets in redshift}. The shape of the entire distribution is dominated by SDSS and DESI.}
    \label{fig:combined_spectroscopicdist}
\end{figure}

\begin{figure}[!ht]
    \centering
    \includegraphics[scale=0.5]{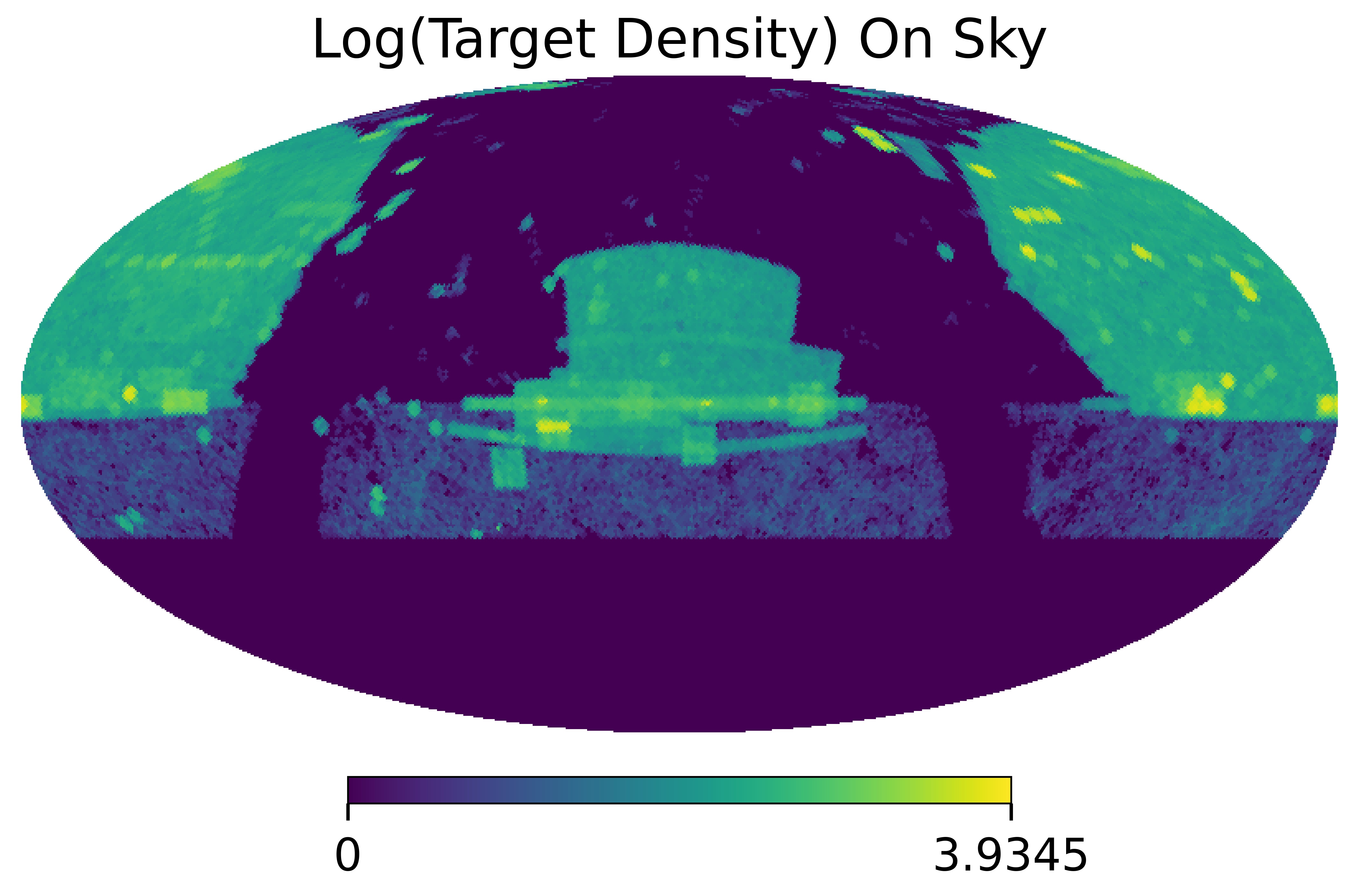}
    \caption{\small\textbf{Distribution of targets In the sky}. Plot as astrometric position on the night sky (RA, DEC). Dominated by the SDSS and DESI footprint, with some deep drilling wells from spectroscopic surveys showing as particularly bright plots. The southern sky south of DEC=$-30^\circ$ is excluded due to the PanSTARRS footprint. The sparse region of the northern sky spans the milky-way.}
    \label{fig:sky_pos}
\end{figure}

\subsection{SDSS}
The vast majority of SDSS spectroscopy comes from three surveys with different targeting guidelines and science goals. The Legacy survey or main galactic survey's goal was to provide uniform spectroscopy and ugriz photometry of extragalactic targets in the northern galactic cap \citep{SDSS_DR8}. It targets galaxies brighter than r=17.77 (80\% of the survey), Luminous Red Galaxies (LRGs) \citep{SDSS_LRG_definition}, and Quasi-stellar objects (QSOs). The main sample of galaxies brighter than r=17.77 is a flux limited sample \citep{SDSS_MGS}, while the LRGs are a volume limited sample designed to probe a specific sub-population of galaxies \citep{SDSS_LRG_definition}, and have a slightly more complicated targeting criteria. There is a strong correlation between satisfying LRG criteria and being an ``early-type'' or Elliptical galaxy.

The BOSS \citep{BOSS_target_guidelines} survey's goals were to vastly expand the number of available galaxies with spectroscopic redshift to perform baryonic acoustic oscillation analysis \citep{BAO}. At low redshifts, BOSS primarily targets LRGs for their strong spectral features allowing precise redshift measurement without long observation time, and beyond redshifts of 0.6 targets a stellar-mass limited sample \citep{BOSS_target_guidelines}. Finally, the extended BOSS (eBOSS) survey \citep{EBOSS} targets LRGs focused at higher redshifts than achieved with BOSS. The effect of the SDSS targeting from BOSS and eBOSS are a sample bias towards the most luminous, red galaxies. While this makes up the vast majority of the sample, SDSS also makes public auxillary programs and surveys completed at a much smaller scale. Nonetheless, it is believed that the sample bias severely restricts empirical photometric redshift algorithms in feature space outside of the flux-limited MGS.

We queried SDSS CAS jobs for all objects in SDSS north of dec>-30$^\circ$ in increments of a few ten degrees.  We accept redshifts that have no bits set on ZWARNING, meaning nothing is ``wrong'' with the redshift measurement. We also make a cut on the respective spectroscopic class being GALAXY. We also query z\_noqso and z\_person, which are additional fields in SpecObjALL. If z\_person is set, we accept z\_person instead. If Z\_noqso is set, we accept Z\_noqso instead. If neither Z\_person or Z\_noqso is set (majority of the sample) we accept the standard reported redshift of the target.  

\subsection{DESI}
DESI targets four sub-populations of galaxies, with which by our cut to ``galaxies'' we are concerned with the bright galaxy sample (BGS), the LRG sample, and the emission line galaxy (ELG) sample. We document the kinds of galaxies and at what redshifts DESI targets them below, and then describe the sample included in the DESI early data release used in this article. The BGS will include a magnitude-limited sample to r<19.5, which will extend our representative sample of galaxies to redshifts of approximately Z=0.6 \citep{DESI_BGS}. The LRG sample builds upon the works of the SDSS surveys, targeting LRGs in the redshift region of 0.4<Z<0.8 in a higher density than SDSS \citep{DESI_LRG}. The ELG survey, traces the highest regime of star-forming galaxies at the peak of the highest star-forming times in the Universe, and will observe targets in 0.6<z<1.6, specifically focusing on the region 1.1<z<1.6 \citep{DESI_ELG}. 

These surveys are on-going and at time of writing only the DESI early data release (EDR) was publicly available \citep{DESI_EDAspectroscopy}. The DESI EDR was used as a science validation sample for the targeting and quality of the DESI surveys, and includes 430,000 bright galaxies, 230,000 LRGs, and 440,000 ELGs.

We downloaded all the spectroscopic targets from the early data release of the DESI \citep{DESI_EDAspectroscopy}. We made quality cuts on ZWARN=0, and spectroscopically confirmed as a galaxy. We also cut at the PanSTARRS field so only accept targets above -30$\circ$. Finally we drop all second or higher references to the same photometric object (as given by photoObjID\_survey).

\subsubsection{DEEP2}
The DEEP2 survey was designed to probe galaxy evolution and the large scale structure with sufficient high quality redshift determinations specifically for fainter objects at greater redshifts \citep{DEEP2_design}. It forms a magnitude-limited sample above redshifts of 0.75.
We download the entire DEEP2 \cite{DEEP2} catalogue and make cuts on declination greater than or equal to -30$\circ$, type of best-fitting template = galaxy, and redshift quality code either secure or very secure. We record zBest for the redshift of the remaining set of galaxies.

\subsection{GAMA}
The GAMA survey target galaxies to r-band magnitude  16.6<r<19.8 with aims to study the dark matter halo mass function, the star formation efficiency of galaxies, and to make a measurement of the recent galaxy merger rate \cite{GAMA_DR2}.
We download all the spectroscopic targets (all galaxies for GAMA) and accept objects with photo-z quality=3 or quality=4 (which should amount to a 95\% probability or confidence in the photo-z). We drop objects with repeated CATID, and filter out objects with a declination less than $30^\circ$.

\subsubsection{VVDS}
The VVDS survey's goals were to compliment the SDSS and 2dF \citep{2dF_finalrelease} survey probing galaxy evolution by surveying galaxies at higher redshifts than those works with a simple selection function that $I_{AB}$<24mag.  
We download the final data release of the VIMOS VLT Deep Survey (VVDS) \citep{VVDS_final_release} which includes the catalogues of the individual WIDE, DEEP and Ultra-DEEP VIMOS surveys. We make a cut on the zflag table to values of 3 or 4. 

\subsection{VIPERS}
The goal of the VIPERS survey \citep{VIPERS_final} was to construct a complimentary survey at greater depths than SDSS, but over a large volume of sky. The selection functions were designed to focus the survey on the redshifts greater than 0.5 and to a depth of $i_{AB}$ < 22.5 mag. 
We download the entire VIPERS catalogue and make cuts on declineation greater than or equal to -30$^\circ$, that the reported zflg quality column indicates either a high confidence, highly secure, or very secure (indicating at least an estimated 90\% confidence in the redshift values). We make a cut on classFlag to only accept main VIPERS galaxy targets.

\subsection{6DF}
The goal of the 6dF \cite{6dFGS_final} survey was to provide a survey of nearby galaxies over a larger volume of the sky than contemporarily available in the preceding 2dF \citep{2dF_finalrelease} or SDSS surveys. The survey centers at a median redshift of 0.05 with a designed K-band limit 12.65. 
We downloaded the 6dF Galaxy Survey Redshift Catalogue Data Release 3, which contains spectroscopy, redshifts, and peculiar velocities for a selected sample of near-infrared observations primarily compiled from the 2MASS catalogue. The 6dF catalogue includes observations from other surveys including SDSS, 2dFGRS, and ZCAT that satisfy their target constraints. We make a cut on the quality column and accept values of 4 or 3.

\subsection{WiggleZ}
The WiggleZ survey was designed to probe large scale structure and dark energy \cite{WiggleZ_final}. The sample is selected by UV brightness using the GALEX space telescope (NUV < 22.8 mag). Additional targeting criteria were used to specifically target ELG. We download the complete WiggleZ catalogue and make cuts on the reported quality of the redshift measurements, QUALITY = 4 or QUALITY=5, 5 meaning an excellent redshift with high S/N that may be suitable as a template. 4 meaning a redshift that has multiple (obvious) emission lines all in agreement. We also make our cut on the declination of the target being greater than or equal to -30$^\circ$.

\section{Photometric Data Collection}
\label{appendix:photometry_download_and_description}

\begin{table}[!ht]
\caption{\small\textbf{Photometric Band Table} Summarizes the properties of the photometric sources used in this paper. Depths are listed as astronomical magnitudes, but the number of objects detected in any particular survey depends on the intrinsic population luminosity of objects in each band. E.g., by visual inspection, many GALEX pointing do not show a signature of the target galaxy.}
  \centering
 \label{tab:photo_filter_table}
  \resizebox{\textwidth}{!}{\begin{tabular}{lllll}
    \toprule
    Source & Ref & Filters & Depth & Pixel Scale\\
 \hline
     GALEX & \citep{GalexSurveyPaper} & FUV, NUV & 19.9, 20.8 & 1.5" \\
     PanSTARRS & \citep{PSSurveyPaper} & g, r, i, z, y & 23.3, 23.2, 23.1, 22.3, 21.3 & 0.25" \\
     UnWISE & \citep{UnWISE_og} & W1, W2 & 20.72, 19.97 & 2.75" \\
    \bottomrule
  \end{tabular}}
\end{table}

We combine photometric cutout data collected from GALEX \citep{GalexSurveyPaper}, PanSTARRS \citep{PSSurveyPaper}, and WISE \citep{WISE_og} observations (the latter re-processed by UnWISE \citep{UnWISE_og}, for filters $W1$, $W2$) for a total 9 different bands of light. The Photometric bands are summarized in table \ref{tab:photo_filter_table}. On a basic level, we chose the PanSTARRS survey for its depth and large footprint, then chose GALEX and UnWISE observations for their nearly all sky completion at relatively high depth. We are notably missing any NIR wavelengths, (e.g., that could be provided by 2MASS \citep{2MASS}, UKIDSS \citep{UKIDSS}, VISTA \citep{VISTA_overview_telescope}); specifically, 2MASS data is comparatively shallow compared to the depth of our targets, while UKIDSS and VISTA are not an all sky survey so a relatively small number of our spectroscopic targets would have available data. In this work we are interested in explaining the behavior of a model that could reasonably be deployed on a wide swath of the night sky. Planned upcoming surveys in the NIR (Euclid \citep{eulid} and SPHEREx \citep{SphereX}) would make excellent additional photometric inputs to our work. %We describe these missions in detail below.

% \subsection{Galex}
% \textcolor{red}{}

% \subsection{Panstarrs}
% \textcolor{red}{}

% \subsection{UnWISE}
% \textcolor{red}{}

\section{Photometry Preprocessing}
\label{appendix:photometry_processing}
\subsection{Photometric Data Processing}
%Downloading the photometric data from cutout tools
Photometric Data can be accessed from server cutout API tools\footnote{GALEX Cutouts: www.legacysurvey.org/viewer/fits-cutout?ra=RA\&dec=DEC\&size=32\&\\pixscale=1.5\&layer=galex}\footnote{PanSTARRS Cutouts: https://ps1images.stsci.edu/cgi-bin/fitscut.cgi?size=170\&format=fits\\\&ra=RA\&dec=DEC\&red=/rings.v3.skycell/1063/090/rings.v3.skycell.1063.090.stk.g.unconv.fits}\footnote{UnWISE Cutouts: www.legacysurvey.org/viewer/fits-cutout?ra=RA\&dec=DEC\&size=32\&\\pixscale=2.75\&layer=unwise-neo7}. We query public APIs of image cutout servers from each photometric survey source at the target coordinates. This process can be easily automated with standard wget tools, but data collection is throttled by server response times and one must take care not to overload servers with requests. Data collection of our entire dataset takes 35 days, for a total volume of approximately 1.5 TB for 4.2e6 samples after quality cuts and various losses. In this preliminary report, we limit our sample to the range $y_i \in [0,1.0]$ and we randomly sample to 7\% of this total sample, $N_{\text{eff}}$ = 2e5 samples. This ensures we can load our sample onto volatile memory at the start of training to avoid IO bottleneck of loading individual batches throughout training. This choice allows us to be more agile in this preliminary investigation. Downloaded files are in the FITS \citep{FITS} file standard. Images are downloaded so that a 30 arcsecond per-side cutout can be created from rotated images without interpolation. We downloaded PanSTARRS at 170pix/side, GALEX at 32pix/side, and UnWISE at 32pix/side. We use UnWISE DR7 \citep{7_yr_UnWISE}.

%processing the photometric data-- dealing with NaNs
After downloading the cutout files we inspect them for masked pixel values. The PanSTARRS data has masked pixels with mask arrays that in principle could be downloaded to explain why some pixels are masked. Because these missing pixels often do not overlap with the host galaxy's profile (in our visual inspection), we believe it is appropriate to in-paint to fill in these masked regions. We filter out PanSTARS images that have more than 1\% NaN pixels in any band, otherwise, we interpolate with cv2\footnote{https://docs.opencv.org/3.4/df/d3d/tutorial\_py\_inpainting.html}. This pre-processing step allows us to keep images containing NaNs from the cutouts inside our input volume. GALEX and UnWISE were not observed to contain any NaN pixels, though GALEX does contain 0-valued pixels wherever a patch of sky is queried where observations were not taken.

%How is the data stored on the data archive I will be releasing.
Images are saved into individual files in a large directory structure. Filenames are set as the $\texttt{Mantis-Shrimp}$ ID. It is necessary to save the file individually so that our data-loader can load file individually to achieve random sampling. This is not ideal, as modern HPC computers including our local cluster use distributed file systems (e.g., Lustre, Qumulo), which dislike large directories of many files and IO suffers from loading / transferring many files. Each file is saved as a python pickle dictionary containing numpy arrays of the PanSTARRS, GALEX, and UnWISE photometry. In this preliminary report we re-save a fraction of our total data into a single file that is stored contiguously in memory.

%Final Processing Steps
%\subsection{Final Processing Steps and Data Loading}
Finally, our images are scaled using the arcsinh scaling (commonly referred to as luptitudes \citep{Luptitude}). Arcsinh scaling is critical as in the linear flux-scale the data has already been centered so some pixels initially have negative values.  We use a softening parameter $\alpha$ = 0.2. for GALEX and UnWISE and $\alpha$ = 0.1 for PanSTARRS.

The extinction due to dust along the line of sight is included as an additional input to the neural network. We utilize both the corrected SFD map \citep{CSFDMap,SFDMap} and the Planck16 map \citep{Planck16_dustmap}. Extinctions are queried for each object using the DUSTMAP\footnote{https://dustmaps.readthedocs.io/en/latest/index.html} python library \cite{DUSTMAPs}.

%Augmentation
\subsection{Augmentation Pipeline}
% We describe two different augmentation pipelines depending on whether we are evaluating early or late-time fusion models. These augmentation pipelines differ only in the final step.
Data augmentation is observed to increase generalization and reduce variance over the transformations trained over.
We perform random rotations and random flips of the incoming data. To avoid interpolation of data due to missing values under rotation, we queried for additional pixels for each cutout that we then crop-out. %For the early time fusion models 
We ensure that the images are placed onto the same size pixel grid for this early fusion experiment. We accomplish this by up-sampling the GALEX and UnWISE data to a $120 \times 120$ pixel grid, spanning 30 arcsec on each side.

\section{Image Examples}
\label{appendix:image_examples}
\begin{figure}[!ht]
    \centering
    \includegraphics[scale=0.4]{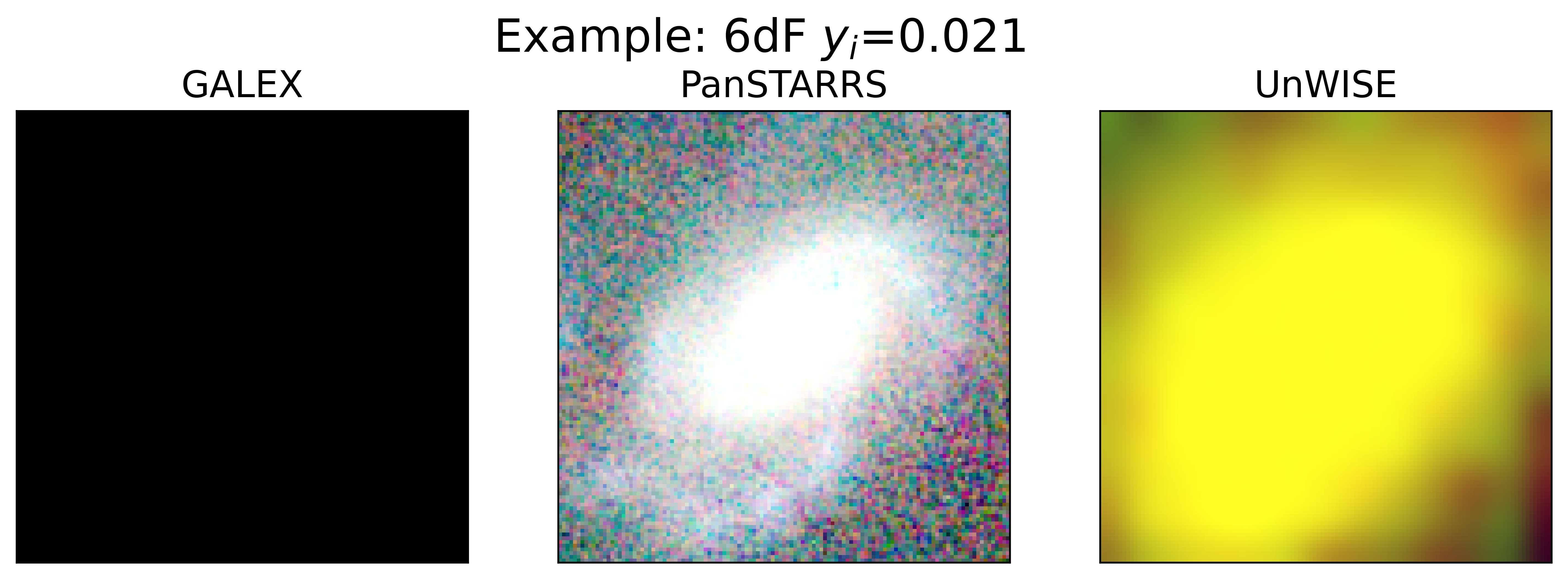}
    \includegraphics[scale=0.4]{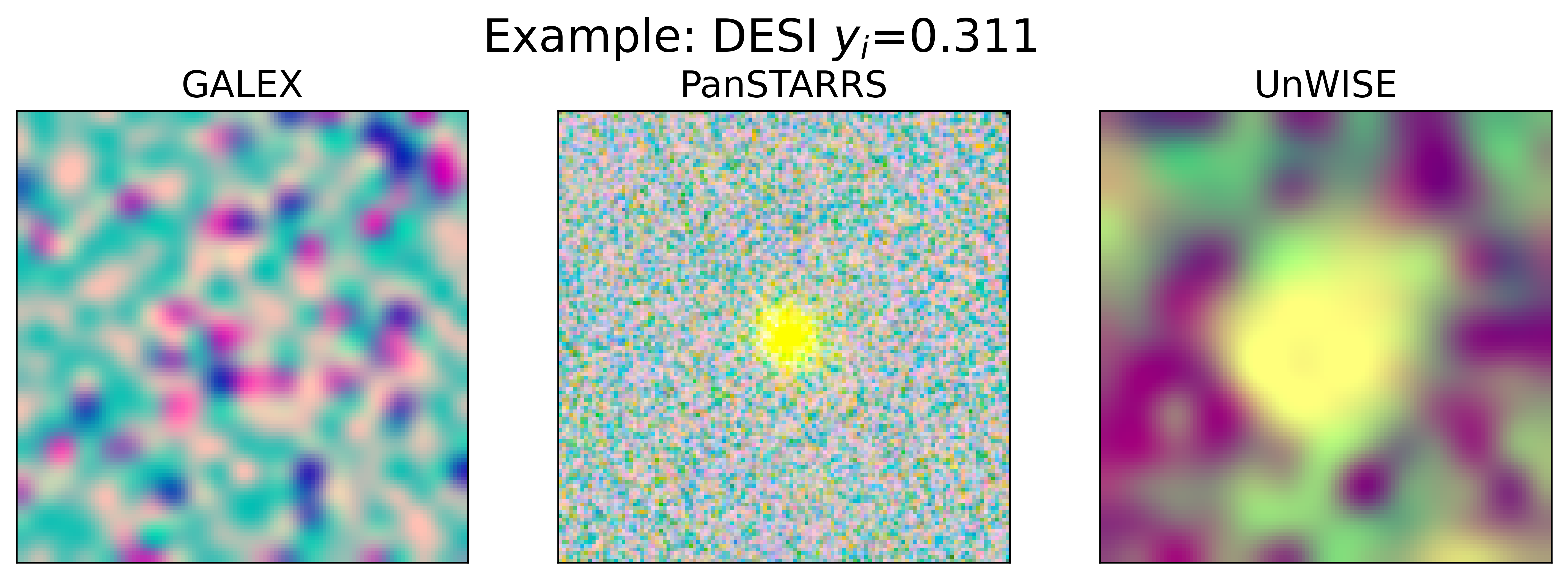}
    \includegraphics[scale=0.4]{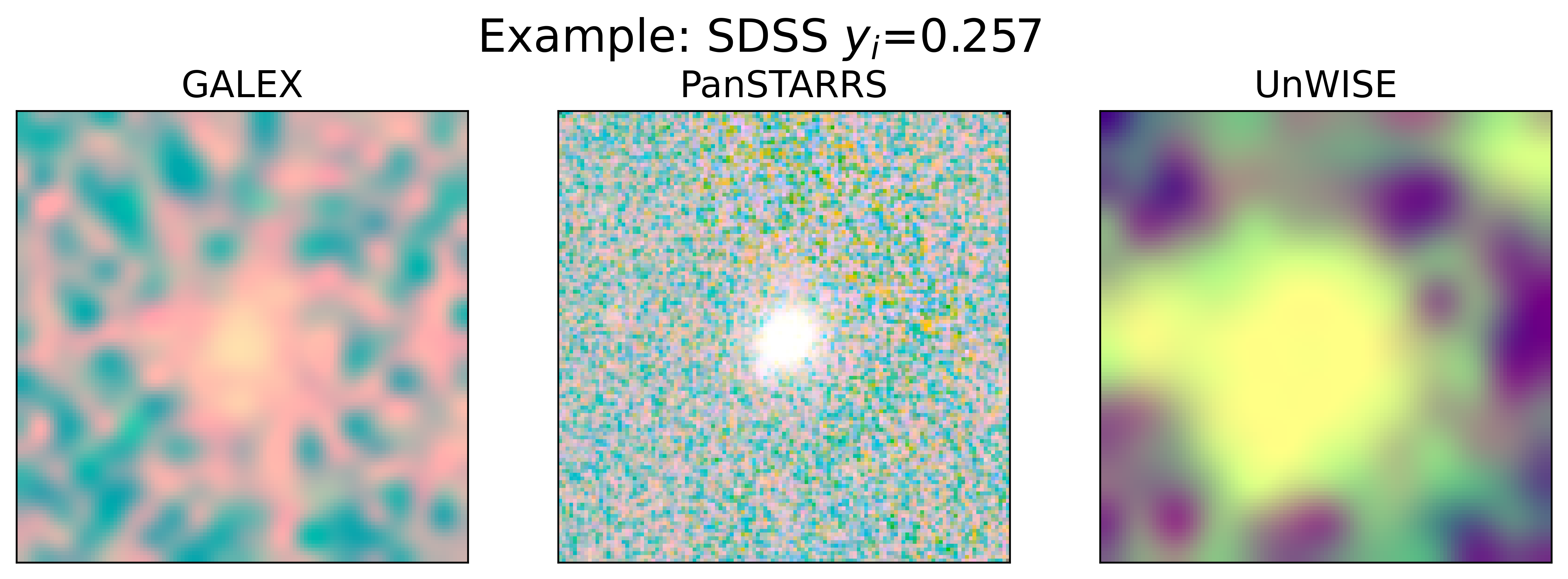}
    \caption{\small\textbf{Examples of each survey in each band} The black GALEX image is caused by missing photometry in the NUV.}
    \label{fig:example_photometry}
\end{figure}

%this appendix just has a few examples of our images with colorbars and gives visualization of the distribution of our images using histograms.
In this appendix we provide three examples of photometry and target labels from our sample, see figure~\ref{fig:example_photometry}.

\section{Shapley Values}
\label{appendix:shapley}

Shapley values \citep{ShapleyOG} have seen a resurgence for the application of neural network explainability \citep{ShapleyRepopularizedforNN}. Shapley values measure how ``valuable'' a player is to a team in an N-person game. These values answer how to fairly distribute rewards to players based upon their synergy given all other players. Let F be the total number of channels, and each channel be $f_j$. We create subsets $S \subseteq \{1,\ldots,n\}$, where $val(S)$ is the model's point prediction of redshift on the modified $x'_i$ input where if $j \notin S$, then we replace $f_j$ with the baseline channel $f'_j$, representing no feature contribution from that channel. The Shapely value for filter $f_j$ is:
\begin{equation}
  \phi_j = \sum_{S \in \{1,\ldots,n\}/\{j\}}  \frac{F!}{|S|!(F-|S|-1)!} \times (\text{val}(S \cup {j}) - \text{val}(S)).
\end{equation}
%MM-SHAP = Multimodal Shapley values.
We note that $\phi_j$ can be positive, negative, or zero indicating whether the channel is increasing (positive) or decreasing (negative) the estimate of redshift. However, we seek to answer the relative importance of channel $f_j$, therefore we employ the MM-SHAP \citep{WhichInputMultiModal}, such that we define the the final contribution as the normalization of the absolute magnitude of the Shapley value: $\text{MM-SHAP}(f_j) = \frac{|\phi_j|}{\sum_j^F |\phi_j|}$. Finally, because Shapley values and $\text{MM-SHAP}(f_j)$ are specific to each $x_i$, we will monitor the distribution of $\text{MM-SHAP}(f_j)$ over bins of true redshift.

\section{Additional Architecture and Training Details}
\label{appendix:training}
We train a ResNet50 \citep{ResNetHe2015} CNN modified to accept input images with 9 channels in the PyTorch framework \citep{PyTorchPaszke2019}. Convolutional and Linear layer weights are initialized from the He Normal distribution \citep{HeInitialization_andPReLU}; bias layers are initialized to zero, and BatchNorm \citep{BatchNormOG} weights are initialized to unity. We use the Leaky ReLU \citep{LeakyReLU} activation function throughout the neural network. The cross entropy loss is minimized with the Adam optimizer \citep{AdamOptimizerKingma2014} with Nesterov momentum incorporated\citep{NAdam} (i.e., NAdam) with default hyperparameter values, except for the initial learning rate which was set to 5e-3. We anneal the learning rate throughout training using a learning rate scheduler tracking the validation cross entropy loss, which reduces the effective learning rate as $\eta_{\text{new}} = \eta_0 * 0.5$, if the validation loss is not seen to improve over a period of six epochs. We use a batch size of 32. We sample from our training dataset using a weighted random sampler with weights set to provide a uniform sampling over the class occurrences. Because we sample from the training data with replacement, its possible to see the same sample more than once per epoch and we set a static 5172 batches per epoch (the same number of batches to use all our training dataset under non-random sampling). We train for 85 epochs, or about 10 hours of training on a NVIDIA V100 GPU.  We track both the loss and our evaluation metrics at regular intervals on the validation set. All hyperparameter tuning (performed by hand in this preliminary investigation) was conducted on this same validation set. Only once was the test set used to evaluate and compute the metrics reported in our main results.

\begin{figure}[!ht]
    \centering
    \includegraphics[scale=0.5]{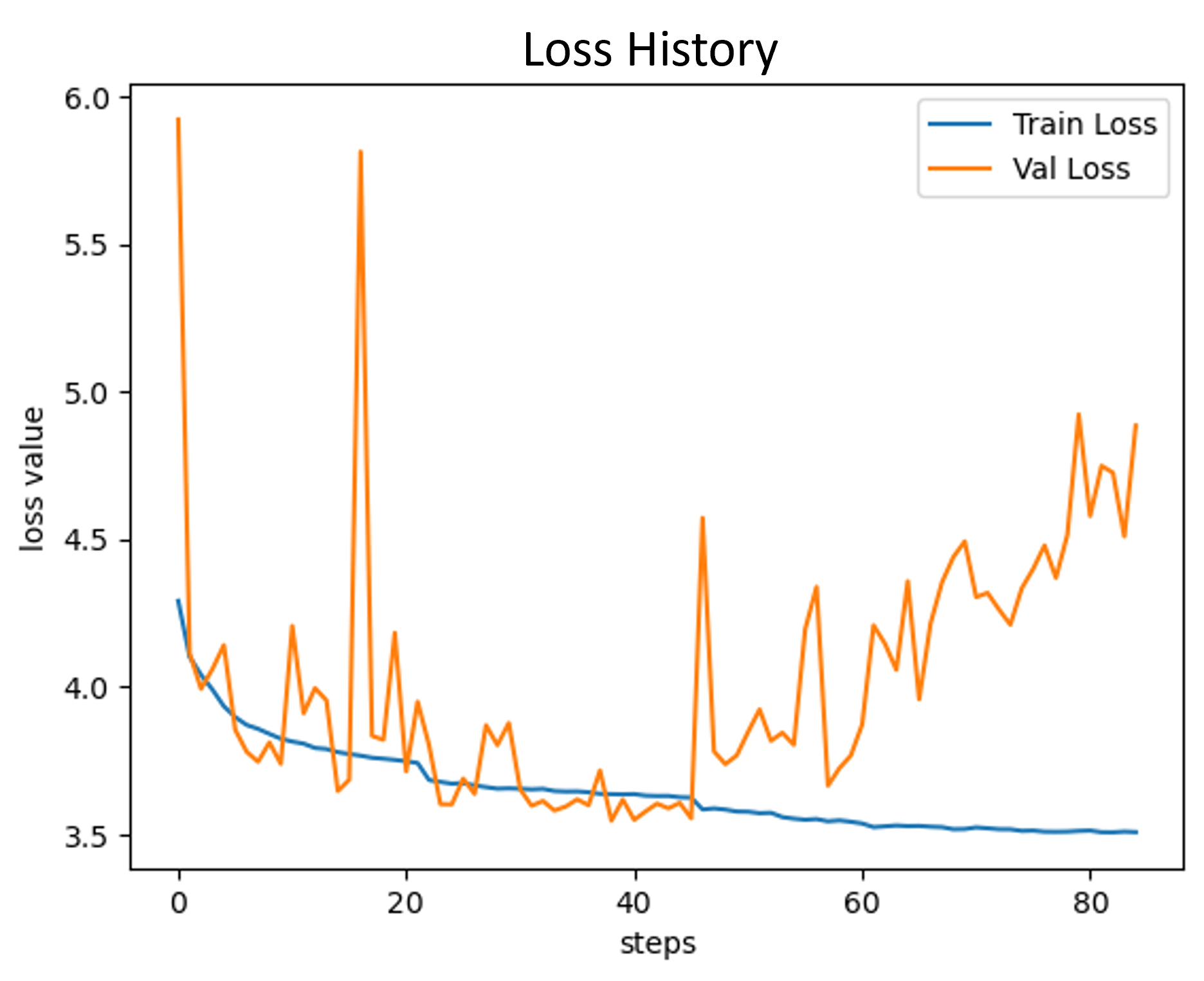}
    \caption{\small\textbf{Loss History} After each epoch we compute the sample-averaged cross entropy loss evaluated on our entire training set and validation set.}
    \label{fig:loss_history}
\end{figure}

\begin{figure}[!ht]
    \centering
    \includegraphics[scale=0.5]{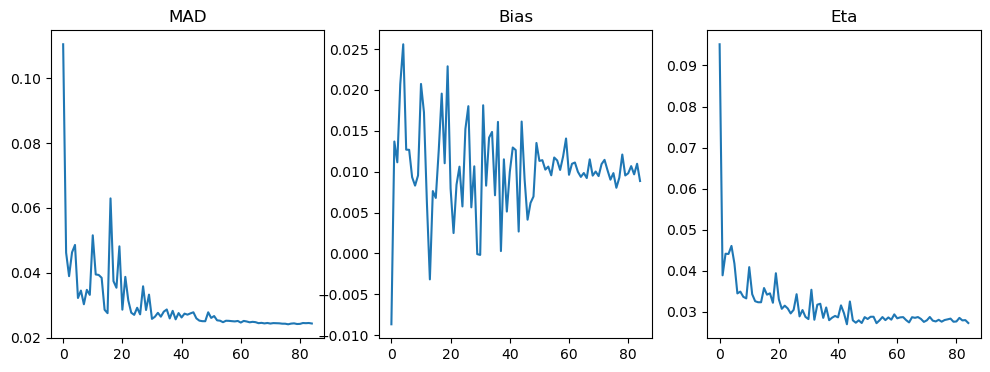}
    \caption{\small\textbf{Metrics over training epochs.} We monitor the metrics of our point estimate over training, which generally trend with the validation loss decreasing, up until epoch 44. Interestingly, while our model appears to over-fit when visualized by its loss increasing, the metrics no longer trend with this increase in loss.}
    \label{fig:metrics}
\end{figure}

We provide the training loss history in figure \ref{fig:loss_history}, which shows that after epoch 44 the validation loss spikes, probably at the same time there is a sharp decrease in the training loss from our LR scheduler. Despite the general trend towards higher losses after this epoch, the underlying point-estimate metrics remain stable, perhaps even losing epoch-wise variance seen earlier in training. The metrics are visualized in figure~\ref{fig:metrics}.

\section{Probability Calibration and Interpretation}
\label{appendix:PIT}

\begin{figure}[!ht]
    \centering
    \includegraphics[scale=0.7]{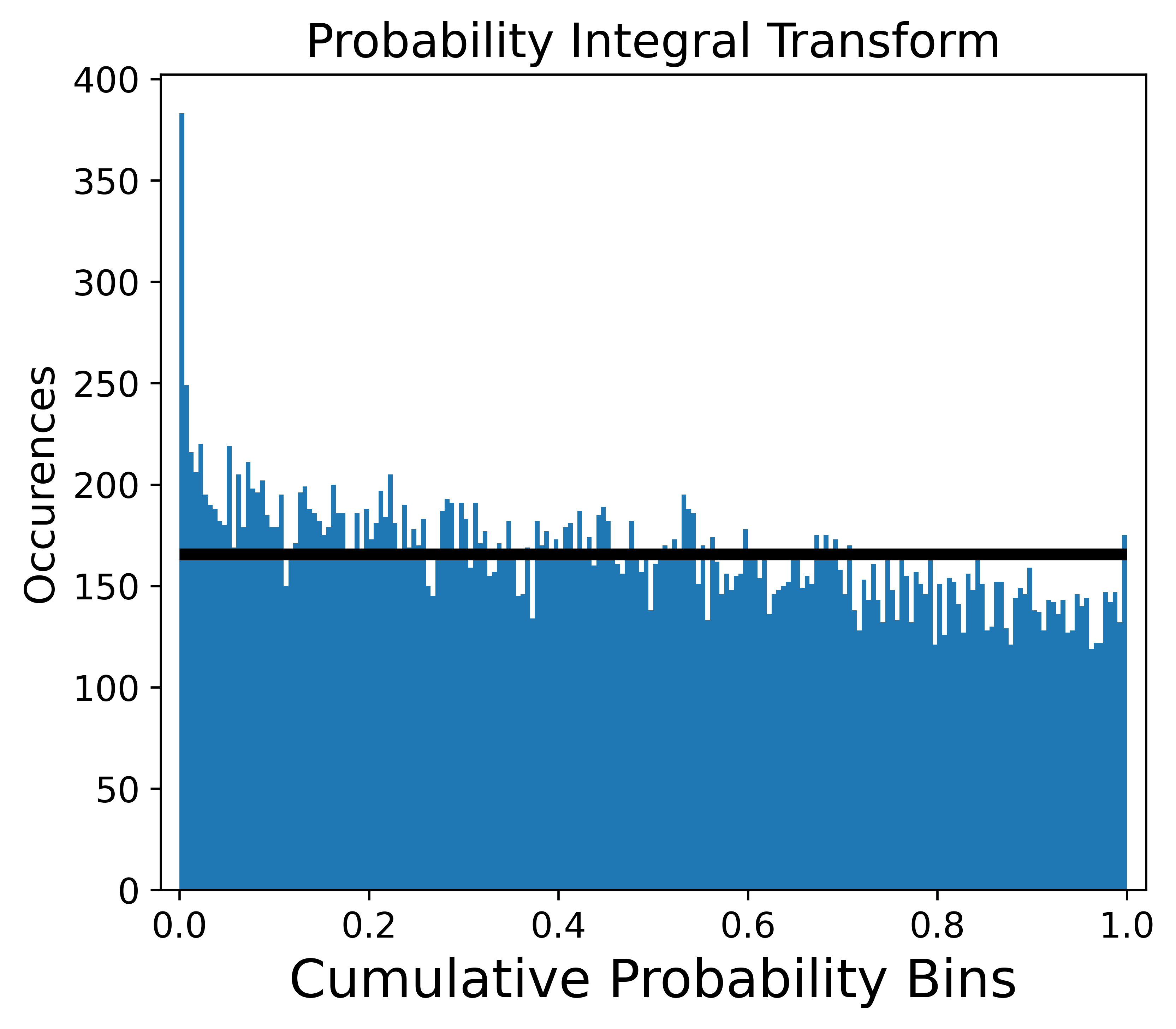}
    \caption{\small\textbf{PIT} The rate of true occurrences in the CDF across the ensemble of test datapoint PDFs generated by our model. The diagonal slope represents that too many true values occur early in the CDFs, representing the large positive bias our network exhibits.}
    \label{fig:PIT}
\end{figure}

The metric of choice for model calibration in photo-z field is the probability integral transform (\textbf{PIT}) \citep{PITOGDawid1984,PITAstroPolsterer2016}, which has been used to compare probabilistic performance across photometric redshift models \citep{PZEvaluationLSSTSchmidt2020}. The PIT is a histogram of the occurrences of true redshift in the set of CDFs measured from density estimates $z_i$, or $\text{CDF}(z_i,y_i) = \int_0^{y_i} z_i \mathrm{d}z$. Essentially, for a perfectly calibrated model each histogram bin would line up with the black horizontal line in figure~\ref{fig:PIT}. A general diagonal downward trend indicates a positive bias, as more true values of redshift occur early in the CDF than a true representation of probability would represent. Its important to remember that PIT is evaluated on the entire ensemble of our test dataset and does not guarantee the probabilistic calibration of any individual estimation, which is an open problem in uncertainty quantification for SciML.

\begin{figure}[!ht]
    \centering
    \includegraphics[scale=0.4]{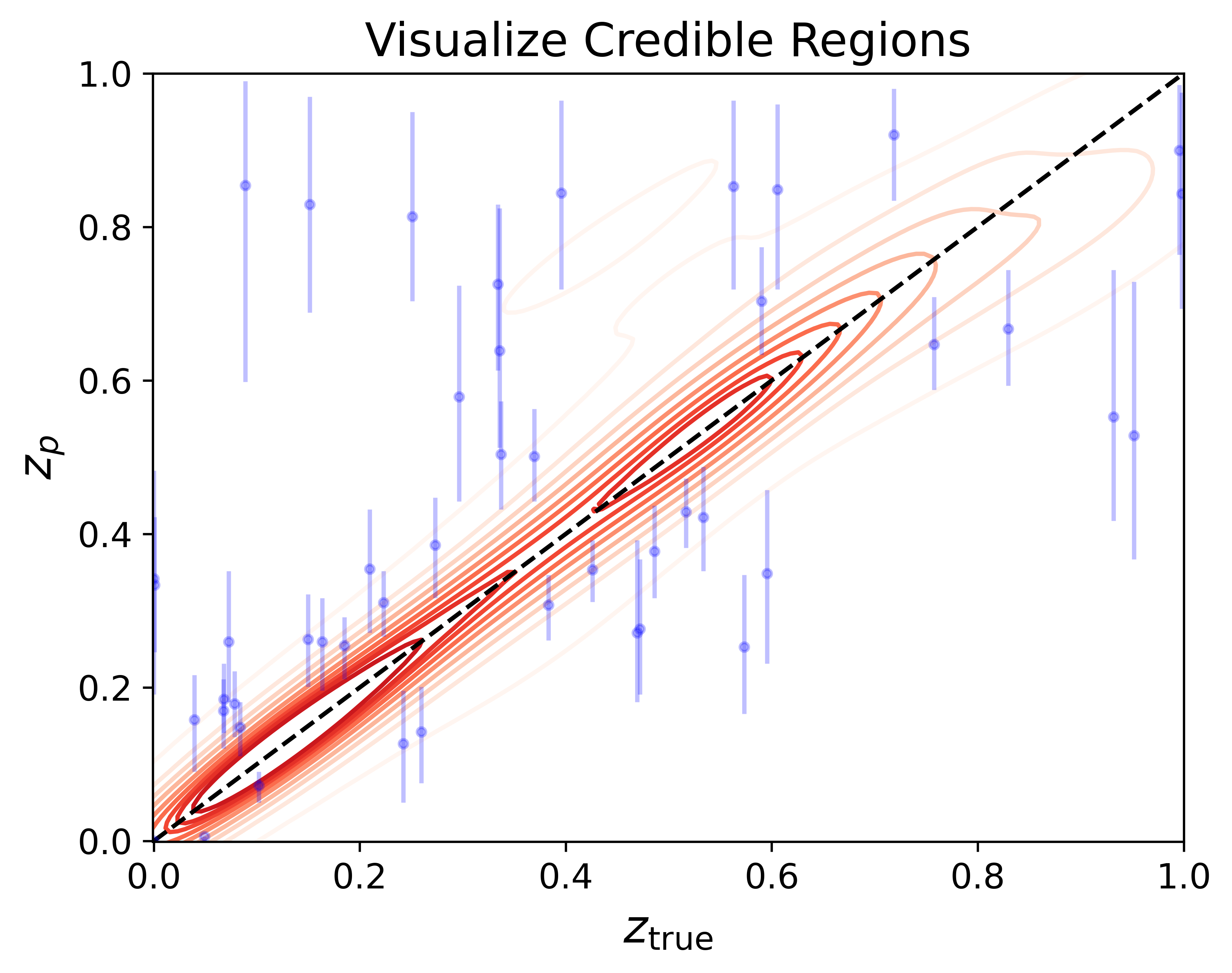}
    \includegraphics[scale=0.4]{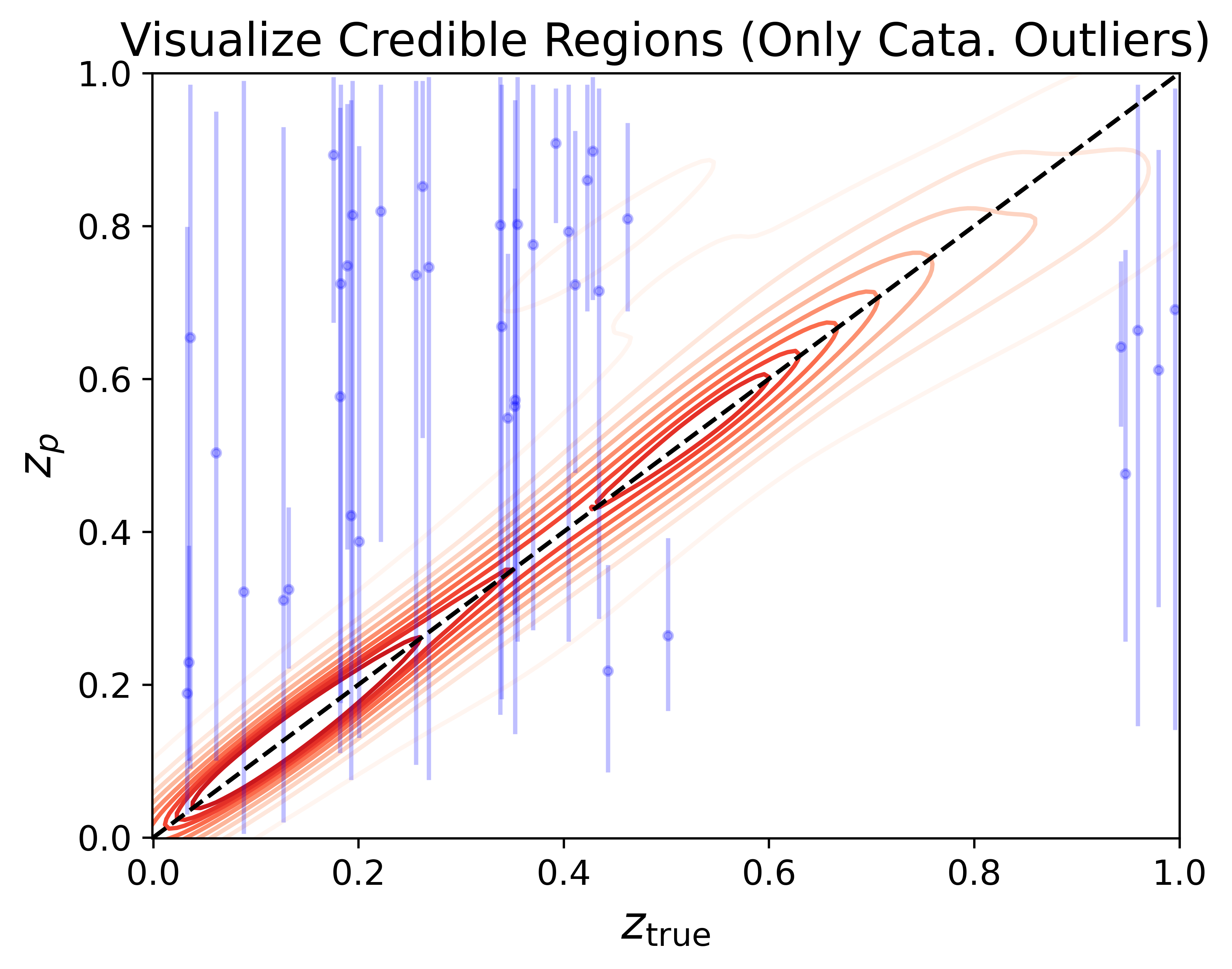}
    \caption{\small\textbf{point-to-point clouds} Both plots show the point-estimates (y-axis) against the ground truth redshift values (x-axis) visualized as a kernel density plot. The Blue ponts with errorbars are a random selection from the population of (Left) points outside the 90\% confidence region estimated from our PDFs and (Right) points classified as ``catastrophic errors'' with scaled residuals greater than 0.05. }
    \label{fig:p2p}
\end{figure}

We can demonstrate how our probabilistic interpretation of the output of our neural network allows us to create confidence regions unique to each point prediction, as in \ref{fig:p2p}, where we show a KDE of the point estimates to true redshift with a small random sample of the test points with error bars demonstrating the 90\% confidence region extracted from the CDF. Our PDF interpretation of the neural network output could be useful, for example, in communicating bi-modality to an end-user.  

\section{Preparing Comparison Datasets}
\label{appendix:comparison_details}
%describe how we query SDSS for SDSS MGS like galaxies

%describe the WISE-PS1-STRM work in enough detail to talk about the kinds of filters being applied when we compare to their work.

%\section{Optional Appendix on What we understand about spectroscopy}
%Deciding whether this appendix helps or not.

In addition to reporting our results on the held-out test dataset, we compare our analysis with the photo-z estimates from two available catalogues, the \citep{Pasquet2019} (P19) catalogue created from the SDSS spectroscopy (including BOSS and EBOSS) of galaxies with a dust corrected SDSS Petrosian magnitude $r$ < 17.77 (mimicking the SDSS main galactic sample's photometric cutoff \citep{SDSS_MGS}), and the B22 catalogue (otherwise known as the WISE-PS1-STRM \citep{WISE-PS1-STRMBeck22} catalogue). 

The P19 catalogue differs from the original in that we query all available spectroscopic targets that satisfy SDSS Petrosian Magnitude $r$ < 17.77 in SDSS DR18 \citep{SDSS_DR18}, the most recent available release at time of writing. P19 was queried from SDSS DR12 \citep{SDSS_DR12}, which was at the time then the most recent available release.  We complete the query on the SDSS CAS jobs server\footnote{https://skyserver.sdss.org/casjobs/}. We simultaneously query for the B16 \citep{Beck2016LLNSDSS} photo-zs calculated using a local linear regression as a second benchmark on this dataset. To form the sample of SDSS MGS we report on, we join the CAS server query results to datapoints that had been randomly sorted into our test dataset from our original train-val-test split. In our test dataset this leaves 7e3 number of samples to evaluate upon. In principle, it would be appropriate to increase this sample by performing a catalog search between all spectroscopic targets from our larger survey in the SDSS footprint that satisfy $r$ < 17.77.

A second relevant dataset that dovetails closely with the main idea of this work, combining multiple surveys together for photo-z information, is \citet{Beck21PS1STRM} and \citet{WISE-PS1-STRMBeck22} (B22), the latter creating the WISE-PS1-STRM catalog. While a full description can be found by reading over these works, briefly: our datasets differ in a few ways. 
First, Beck's work utilizes dense networks on pre-calculated photometric features available in catalogues while our work uses the science images. This generally means that Beck's work relies on a detection in at-least one band of each PanStarrs and WISE for any spectroscopic target. In contrast, our work is more akin to forced photometry centered on spectroscopic astrometry-- we do not require any detections in the underlying photometric survey catalogues. 
Beck also employs a cautious astrometry check between objects in catalogues to en-masse join the catalogues and spectroscopic targets together\citep{CrossMatchBudavariSzalay}. This is entirely circumvented in our methodology since we do not need to query for objects from catalogues, but instead broad patches of the sky from image cutout servers.
B22 does not include the DESI spectroscopic target contribution, as their work was completed before the DESI early release was available to the public. Given that the DESI is nearly a quarter of our sample, we gain a sizable increase in total available targets to train on.
Finally, the WISE-PS1-STRM algorithm actually is two separate AI algorithms: the algorithm performs STAR-GALAXY-QSO filtering, and then on the objects identified as galaxies, will compute a photo-z. This acts as another filter when querying the WISE-PS1-STRM dataset compared to our original training sample.

% \section{Why \textit{Mantis Shrimp?}}
% Ah, the mighty and powerful mantis shrimp. The latin mantis shrimp is known for two cool things: it possesses the most complex eyes, and a whalloping punch. 

\end{document}